\documentclass[10pt,a4paper]{article}
\usepackage{amsmath}
\usepackage{amsfonts}
\usepackage{amssymb}
\usepackage{graphics}
\usepackage{graphicx}
\usepackage{multimedia}
\usepackage{amsmath,amssymb}
\usepackage{epsfig,amsfonts,graphicx,bezier,amssymb}
\usepackage{anysize}
\usepackage{verbatim}

\newtheorem{thm}{Theorem}[subsection]
\newtheorem{lem}{Lemma}[subsection]

\newtheorem{prop}{Proposition}[subsection]

\newcommand{\E}{\mathbb E}

\title{\large {\bf {Modelling Within-Host Immune Response to Visceral Helminthiasis and Malaria Co-infection with Prophylaxis }}}
\author {\bf {\small{B. Nannyonga\footnote{Corresponding author: bnk@math.mak.ac.ug}, J.Y.T. Mugisha~\& L.S. Luboobi}} \\{\small {\it{Department of Mathematics, Makerere University.}
Box 7062, Kampala, Uganda}}}
\begin{document}
\date{}
\maketitle
\thispagestyle{empty}
\begin{quote}{\bf{ABSTRACT}}~~
In this paper, a co-infection of malaria and visceral helminthiasis with immune stimulation and impairment is studied. We have assumed that an individual gets malaria infection during invasion by helminths larvae. In absence of immune response, our results show that the antigens invade the blood system if the rate of red cell rupture per invading merozoite is greater than one.  If fewer merozoites are released, the initial state is globally asymptotically stable. If more merozoites are released, there exists malaria-only endemic point. However, both antigens coexist if the mean infection burden is greater than one. In this case, there is a threshold value for drug action below which no recovery of host is expected. In presence of immune response, three equilibrium states exist. The initial invasion state, and secondly, the unstable state when the immune population has been triggered and the antigens have been eliminated. The third state represents the endemic state which is stable if the infection-induced rupture rate per erythrocyte with immune activation is less than the total mortality of the erythrocytes. A model for severity of the co-infection shows that an immune response will be delayed until immunological barrier values for malaria and helminths are exceeded.
\end{quote}
\section{Introduction}
Specifically, in this model of immune reaction, we explore if an asymptotic decrease of antigen quantity is reached, the cause of periodic course of the illness, and if there is unlimited growth of the antigen quantity within the human, and under what conditions. The model allows a description of threshold relationship between the infection process and the initial dose of the antigen. We seek to determine whether destruction of either malaria parasites or helminths larvae is possible without spread of infection, destruction after either infection, coexistence with specific antibodies, a possibility of recurrent course of infection of either disease, and if there is unlimited multiplication of either malaria parasites, or helminths larvae. It is expected that this co-infection model will give a new insight to researchers in designing experiments for discovering the underlying mechanism of immune response and suppression in within-host malaria and visceral helminthiasis co-infection. We have assumed a normal functioning of the immune system where there is no distinction between the cellular and humoral components of immunity fighting the malaria parasites and helminths larvae that have penetrated into the human blood circulation but that a human being has such defense components, we refer to as antibodies no matter whether it is a cellular-lymphoid system, or  humoral-immunoglobulin system of immunity \cite{CDC}.\\
\section{Model formulation}
The biologically known within-host dynamics of visceral helminths are such that when a mosquito or fly carrying microfilarial larvae bites a new host, it deposits larvae into the subcutaneous tissue of the host. The larvae find their way into the lymph vessels where they take about a year to mature. The mature adult female produces larvae called microfilaria. These escape from the lymphatics to reach blood circulation. It is from here where mosquitoes biting the infected person ingest them \cite{CDC}. The ingested microfilaria develop in the mosquitoes to reach the infective stage of microfilaria and then transmit them to the next human victim when they bite him. This is similar to the transmission of malaria parasites. The injected larvae migrate from blood stream into the lymphatics thus completing the life cycle; when they mature into adult filarial worms. Microfilaria appear in plenty in blood circulation at distinct times of the day, usually in the evenings and at night. This characteristic phenomenon is referred to as periodicity and appears to be related to the biting habits of the transmitting mosquitoes. \\ 

In humans, malaria parasites grow and multiply first in the liver cells and then in the red cells of the blood. In the blood, successive broods of parasites grow inside the red cells and destroy them, releasing daughter parasites (merozoites) that continue the cycle by invading other red cells.\\

The blood stage parasites are those that cause the symptoms of malaria. When certain forms of blood stage parasites (gametocytes) are picked up by a female {\it{Anopheles}} mosquito during a blood meal, they start another different cycle of growth and multiplication in the mosquito. After 10-18 days, \cite{CDC}, the parasites are found (as sporozoites) in the mosquito's salivary glands. When the {\it{Anopheles}} mosquito takes a blood meal on another human, the sporozoites are injected with the mosquito's saliva and start another human infection when they parasitize the liver cells. Thus, the mosquito carries the disease from one human to another (acting as a vector). Differently from the human host, the mosquito vector does not suffer from the presence of the parasites \cite{CDC}.\\ 

Consider a population of $N_e(t)$ erythrocytes in an individual. Let the uninfected erythrocytes be $U_e(t)$. Assuming an invasion of 
malaria parasites (merozoites) with concentration $M(t)$. At the same time, assume that the individual is attacked by a mass of helminths larvae $H(t)$. Each of these two parasites invade random red blood cells and impregnants them. When the invasion is successful, the uninfected cell becomes infected. Let the body of infected red blood cells be $I_e(t)$. Therefore $N_e(t)=U_e(t)+I_e(t).$  After infection, immune responses against the malaria parasites and helminths larvae are triggered. A wide body of evidence indicates that protective anti-blood stage immunity is dependent upon $CD4^+T$ cells \cite{Suss}. Therefore, effective immunity is considered as a population of activated $T$ cells and assume that the resulting immune response is a direct function of their density. The magnitude of the immune response is proportional to the density of the immune cells. The immune cells augment the clearance of the merozoites, helminths larva, and infected red blood cells from the body. We assume that immune activation is proportional to the density of the infective stages and the precursors are not limiting. Define $T(t)$ as the number of antibodies recruited when their resting precursors contact free merozoites, infected cells and free larvae at net rates $\gamma_{_M}M, \gamma_{_I}I_e,$ and $\gamma_{_H}H$ respectively. Let the rate at which the immune cells expand be $p$. This rate of expansion encapsulates the positive feedback upon the immune system. We further assume that a regulatory negative feedback force operates to suppress immune population growth at a rate proportional to the square of its density, $bT^2$ \cite{Het}. This function implies a regulation of the response at high antigen concentrations, since convex or plateauing relationships between the rate of $T$ cell proliferation and antigen concentration is observed \cite{Lamb,Mat}. Thus, we have clearance rates of free merozoites and larvae due to $B$ cells and macrophages given by $\lambda_M M T$ and $\lambda_H HT$ respectively. The clearance of the infected red blood cells due to $T$ cells is given by $\lambda_I I_eT$. The rate of change of density of immune cells is described by their proliferation and deaths rates. They proliferate in response to contact with free-merozoites, free larvae and infected red blood cells at rates $\gamma_{_M} M,$ $\gamma_{_H} H$ and $\gamma_{_I} I_e$ respectively. We define the natural clearance rate of the antibodies as $\mu_t$. All parameters of the model are positive real numbers. \\

Let the uninfected erythrocytes be recruited from the bone marrow at a per capita rate $\Lambda$. We further define removal of the erythrocytes due to aging as a per capita rate $\mu_e$. Let $\alpha_M$ be the per capita rate of erythrocyte invasion and comprises of the dual probability of contact between an uninfected erythrocyte and a merozoite, and of such a contact resulting in productive invasion. The corresponding invasion term for helminths larvae is $\alpha_H$. The infected red blood cells experience a per capita death rate $\delta_{_M}$ due to infection-induced rupture where $\delta_{_M}$ is large compared to $\mu_{_e}$. Each ruptured infected cell produces an average of $r$ merozoites such that the net rate of merozoite production is $r\delta_{_M} I_e$ which is reduced by drug action effect $\epsilon$. Free merozoites die at natural rate $\mu_{_M}$. If the helminths larvae are not ingested by a feeding fly or mosquito, they die at a per capita rate $\mu_{_H}$. The larvae suffer additional death due to prophylaxis at an average rate $\delta_{_H}$. If we assume that merozoites and larvae invade erythrocytes at random, that is, have no predilection for erythrocytes of a particular age or type \cite{Het}, then the above definitions and assumptions lead to the following set of non-linear ordinary differential equations:

\begin{eqnarray}\begin{array}{lll}\frac{dU_e}{dt}=&\Lambda-\mu_e U_e-\alpha_{_M} U_e-\alpha_{_H }U_e\\\frac{dI_e}{dt}=&\alpha_{_M}U_e+\alpha_{_H }U_e-\mu_e I_e-\delta_{_M}I_e-\lambda_I I_e T\\\frac{dM}{dt}=&r\delta_{_M}(1-\epsilon)I_e-\mu_{_M} M-\alpha_{_M}U_e-\lambda_{_M}MT\\\frac{dH}{dt}=&\Pi-\mu_{_H}H-\delta_{_H}H-\alpha_{_H }U_e-\lambda_{_H}H T\\\frac{dT}{dt}=&\gamma_{_M} M+\gamma_{_I} I_e+\gamma_{_H} H+p T-bT^2-\mu_t T\end{array}{\label{helmal1}}\end{eqnarray}

First we analyze this system for the case when there is no immune response and determine the existence and nature of the stationary states. 

\subsection{Criteria for invasion and persistence in the system without immune response} 

\begin{figure}[ht!]\centering{
        \includegraphics[width=0.40\textwidth]{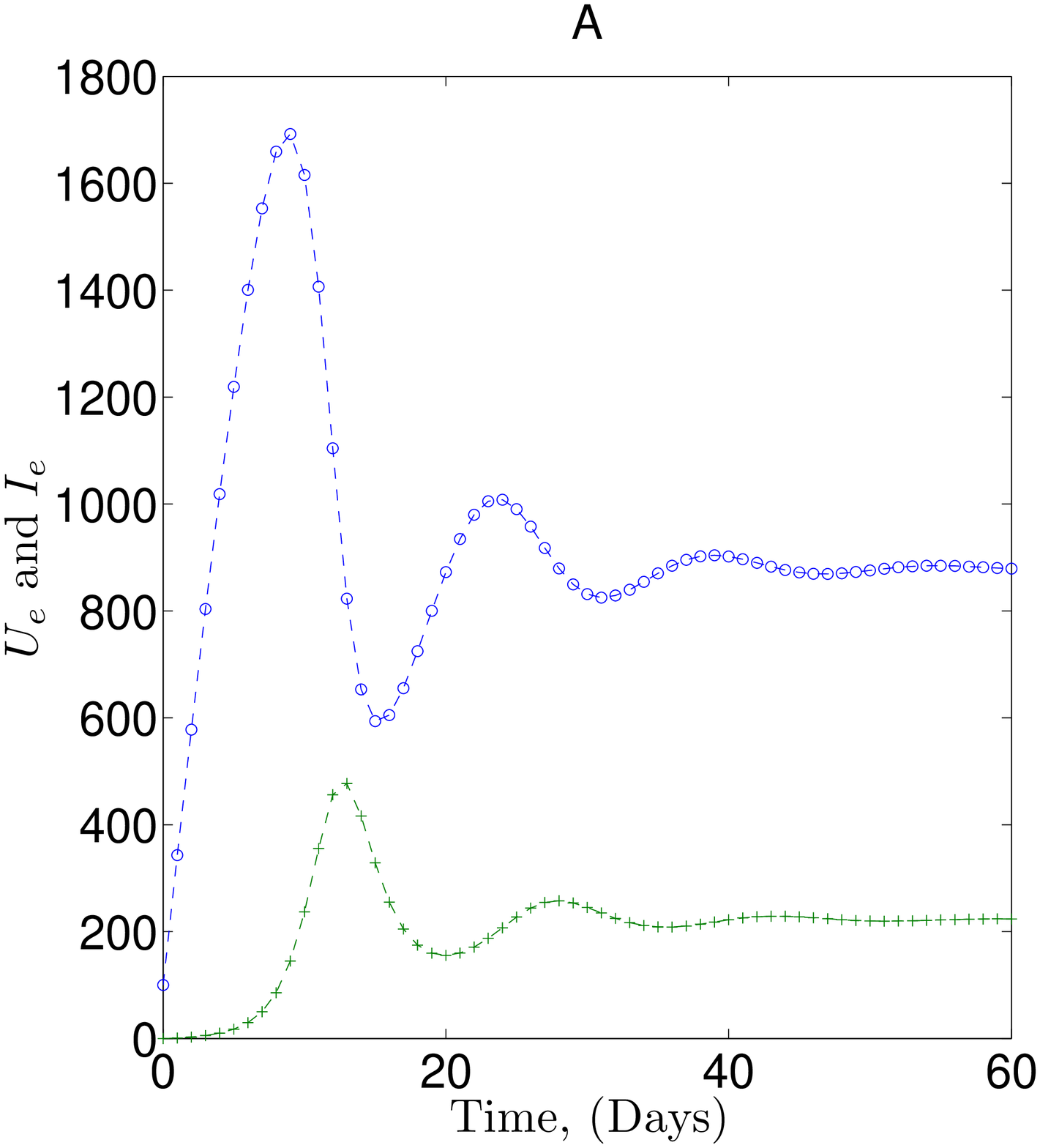} 
       \includegraphics[width=0.40\textwidth]{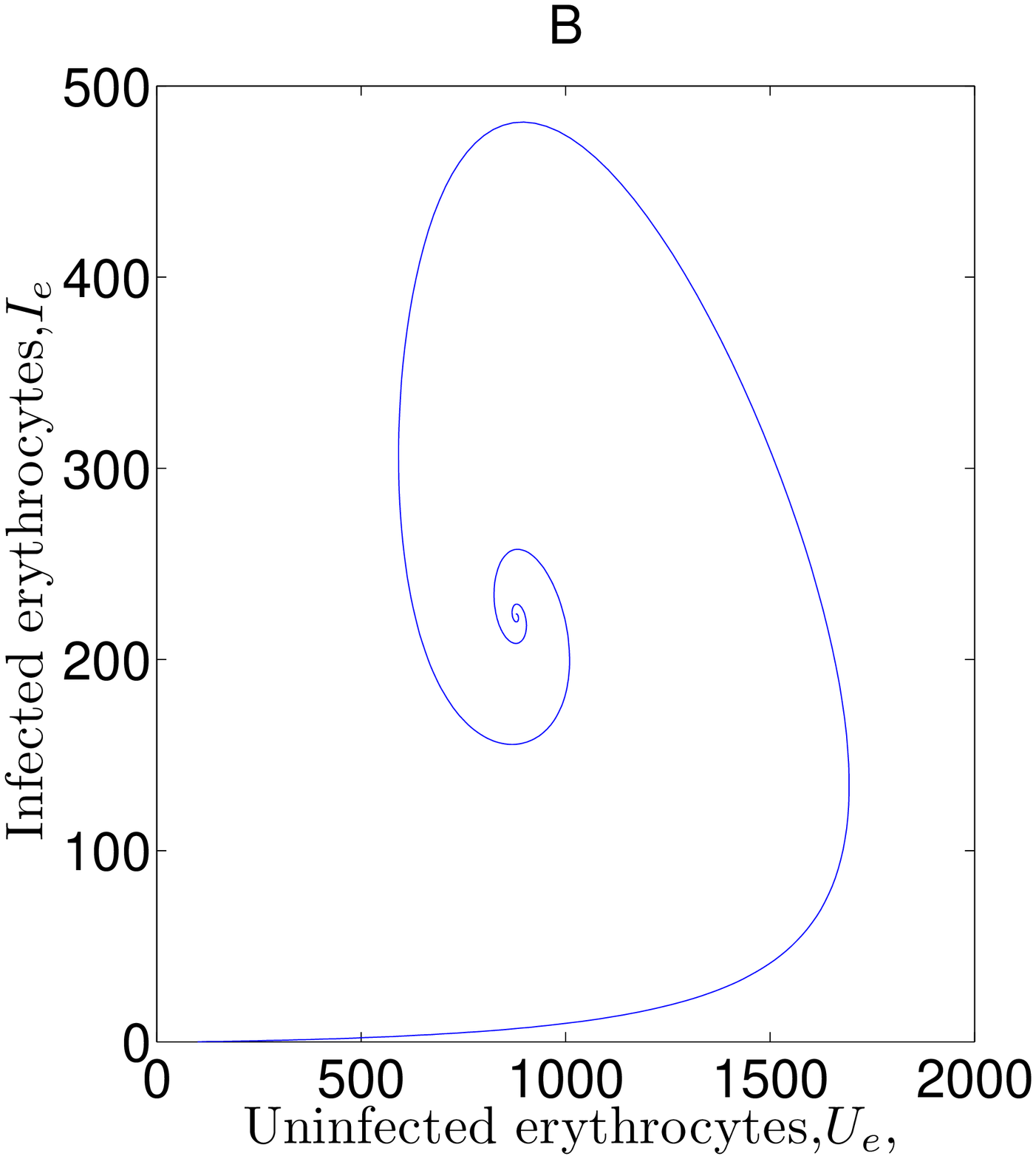}  
       \includegraphics[width=0.40\textwidth]{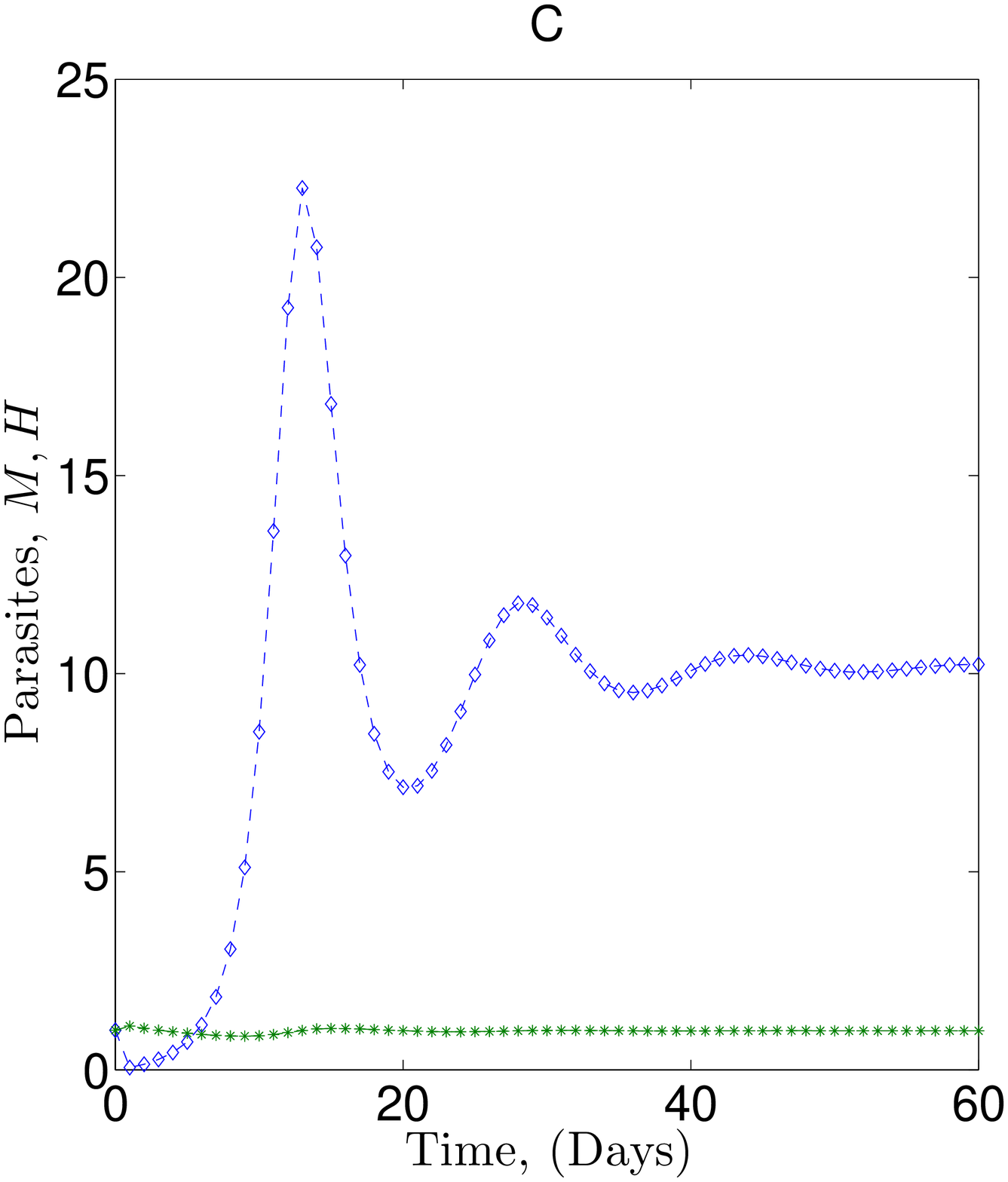}  }
\caption{Graphs for the evolution of uninfected and infected erythrocytes with time (A). In the graph, the circles represent uninfected cells while the crossed line is for infected cells. The phase space portrait in the uninfected $U_e$ and infected $I_e$ space is shown in (B), while the evolution of malaria $M$ and helminth larvae $H$ with time is shown in (C), here, the diamond line is for malaria parasites while the star line represents the helminths larva. $\epsilon=0.5, \Lambda=250, \mu_e0.025, \mu_h=20, \mu_M=48, \delta_M=1, \delta_H=1, \alpha_M=0.025, \alpha_H=0.005, r=16, \Pi=25, \gamma_M=0.1, \gamma_I=0.1, \gamma_H=0.1, p=0.1, b=0.000000001, \mu_t=0.05, \lambda_I=0.00000001, \lambda_M=0.00000001, \lambda_H=0.00000001. $} \label{1}\end{figure}

As in \cite{March}, without immune response System (\ref{helmal1}) simplifies to 
\begin{eqnarray}\begin{array}{ll}\frac{dU_e}{dt}=&\Lambda-\mu_e U_e-\alpha_{_M} U_eM-\alpha_{_H }U_eH\\\frac{dI_e}{dt}=&\alpha_{_M}U_eM+\alpha_{_H }U_eH-\mu_e I_e-\delta_{_M}I_e\\\frac{dM}{dt}=&r\delta_{_M}(1-\epsilon)I_e-\mu_{_M} M-\alpha_{_M}U_eM\\\frac{dH}{dt}=&\Pi-\mu_{_H}H-\delta_{_H}H-\alpha_{_H }U_eH\\\end{array}{\label{helmM2}}\end{eqnarray}
Using the next generation matrix method we show that the basic reproduction number is given by \begin{eqnarray}R_o=\frac{\frac{\Lambda}{\mu_e}\alpha_Mr\delta_M(1-\epsilon)}{(\mu_e+\delta_M)(\mu_M+\frac{\Lambda}{\mu_e}\alpha_M)}
\end{eqnarray} Next we determine the nature of stability of the system. The first steady state is when there is no invasion by malaria parasites given by $U_e^*=\frac{\Lambda}{\mu_e},~I_e^*=M^*=0,~H^*=\Pi$,~ $\Pi \ll 0$. The endemic state is too complicated to obtain useful analytical results. We use numerical simulation to integrate numerically and obtain the solutions as shown in Figure (\ref{1}). These graphs give an example of how endemic stability is attained. The phase portrait indicates that the system moves towards a fixed point at endemic equilibrium.
\begin{lem}
In the absence of immune response the malaria parasites and helminths larvae are only capable of initially invading the blood system if criteria equation $R_o>1$. Biologically, if the rate of red cell rupture per invading merozoite is greater than one.\label{lem1}
\end{lem}
Next we determine the local stability of the equilibrium states. Without immune response, the Jacobian of the system at endemic equilibrium is given by \begin{footnotesize}\begin{eqnarray}\nonumber J(E^*)=\left[\begin{array}{cccc}-\mu_e-\alpha_{_M}M^*-\alpha_{_H }H^*&0&-\alpha_{_M}U_e^*&-\alpha_{_H }U_e^*\\\alpha_{_M}M^*+\alpha_{_H }H^*&-\delta_{_M}-\mu_e&\alpha_{_M}U_e^*&\alpha_{_H }U_e^*\\-\alpha_{_M}M^*&-\alpha_{_M}M^*+r\delta_{_M}(1-\epsilon)&-\mu_{_M}-\alpha_{_M}U_e^*&0\\-\alpha_{_H }H^*&-\alpha_{_H }H^*&0&-\mu_{_H}-\delta_{_H}-\alpha_{_H }U_e^*\end{array}\right]\\{\label{J}}\end{eqnarray}\end{footnotesize}
The malaria invasion-free steady state is stable if 
\begin{eqnarray}R^2_{_{1}}&=\frac{\alpha_{_M}r\delta_{_M}\frac{\Lambda}{\mu_e}(1-\epsilon)}{\left[
(\mu_{_H}+\delta_{_H}+\alpha_{_H }\frac{\Lambda}{\mu_e})(\delta_{_M}+\mu_e+\mu_{_M}+\alpha_{_M}\frac{\Lambda}{\mu_e})+
(\delta_{_M}+\mu_e)(\mu_{_M}+\alpha_{_M}\frac{\Lambda}{\mu_e})\right]}
<1\\ R^2_{_{2}}&= \frac{\alpha_{_M}r\delta_{_M}\frac{\Lambda}{\mu_e}(1-\epsilon)}{(\delta_{_M}+\mu_e)(\mu_{_M}+\alpha_{_M}\frac{\Lambda}{\mu_e})}
<1\end{eqnarray}
{\bf{Proof:}}
We show how the local stability of the initial invasion state without immune response is obtained. The Jacobian of system (\ref{helmM2}) at endemic equilibrium is given by \begin{footnotesize}\begin{eqnarray}\nonumber J(E^*)=\left[\begin{array}{cccc}-\mu_e-\alpha_{_M}M^*-\alpha_{_H }H^*&0&-\alpha_{_M}U_e^*&-\alpha_{_H }U_e^*\\\alpha_{_M}M^*+\alpha_{_H }H^*&-\delta_{_M}-\mu_e&\alpha_{_M}U_e^*&\alpha_{_H }U_e^*\\-\alpha_{_M}M^*&-\alpha_{_M}M^*+r\delta_{_M}(1-\epsilon)&-\mu_{_M}-\alpha_{_M}U_e^*&0\\-\alpha_{_H }H^*&-\alpha_{_H }H^*&0&-\mu_{_H}-\delta_{_H}-\alpha_{_H }U_e^*\end{array}\right]\\\end{eqnarray}\end{footnotesize}
When there is no invasion by either pathogen, $U_e^o=\frac{\Lambda}{\mu_e}$. We therfore obtain the Jacobian at the invasion-free state given by \begin{footnotesize}\begin{eqnarray}\left[\begin{array}{cccc}-\mu_e&0&-\alpha_{_M}\frac{\Lambda}{\mu_e}&-\alpha_{_H }\frac{\Lambda}{\mu_e}\\0&-\delta_{_M}-\mu_e&\alpha_{_M}\frac{\Lambda}{\mu_e}&\alpha_{_H }\frac{\Lambda}{\mu_e}\\0&r\delta_{_M}(1-\epsilon)&-\mu_{_M}-\alpha_{_M}\frac{\Lambda}{\mu_e}&0\\0&0&0&-\mu_{_H}-\delta_{_H}-\alpha_{_H }\frac{\Lambda}{\mu_e}\end{array}\right]\end{eqnarray}\end{footnotesize} with distinct eigenvalue $-\mu_e$. The remaining three are obtained from the 3$\times$3 square matrix given by 
\begin{eqnarray}\left[\begin{array}{ccc}-\delta_{_M}-\mu_e&\alpha_{_M}\frac{\Lambda}{\mu_e}&\alpha_{_H }\frac{\Lambda}{\mu_e}\\r\delta_{_M}(1-\epsilon)&-\mu_{_M}-\alpha_{_M}\frac{\Lambda}{\mu_e}&0\\0&0&-\mu_{_H}-\delta_{_H}-\alpha_{_H }\frac{\Lambda}{\mu_e}\end{array}\right]\end{eqnarray} 
with characteristic polynomial \begin{equation}\lambda^3+a_1\lambda^2+a_2\lambda+a_3=0{\label{E}}\end{equation} where \begin{eqnarray}\begin{array}{ll}
a_1=&\delta_{_M}+\mu_e+\mu_{_M}+\mu_{_H}+\delta_{_H}+(\alpha_{_M}+\alpha_{_H })\frac{\Lambda}{\mu_e}) >0\\
a_2=&\left[(\mu_{_H}+\delta_{_H}+\alpha_{_H }\frac{\Lambda}{\mu_e})(\delta_{_M}+\mu_e+\mu_{_M}+\alpha_{_M}\frac{\Lambda}{\mu_e})+(\delta_{_M}+\mu_e)(\mu_{_M}+\alpha_{_M}\frac{\Lambda}{\mu_e})\right][1-R^2_{_{1}}],\\&\mbox{where $R^2_{_{1}}=\frac{\alpha_{_M}r\delta_{_M}\frac{\Lambda}{\mu_e}(1-\epsilon)}{\left[
(\mu_{_H}+\delta_{_H}+\alpha_{_H }\frac{\Lambda}{\mu_e})(\delta_{_M}+\mu_e+\mu_{_M}+\alpha_{_M}\frac{\Lambda}{\mu_e})+(\delta_{_M}+\mu_e)(\mu_{_M}+\alpha_{_M}\frac{\Lambda}{\mu_e})\right]}$}\\>& 0~~ \mbox{since $R^2_{_{1}} <1$}
\\a_3=&(\mu_{_H}+\delta_{_H}+\alpha_{_H }\frac{\Lambda}{\mu_e})(\delta_{_M}+\mu_e)(\mu_{_M}+\alpha_{_M}\frac{\Lambda}{\mu_e})[1-R^2_{_{2}}],\\
>& 0 \mbox{ if $R^2_{_{2}} < 1$}\end{array}
\end{eqnarray}
where $ R^2_{_{2}}= \frac{\alpha_{_M}r\delta_{_M}\frac{\Lambda}{\mu_e}(1-\epsilon)}{(\delta_{_M}+\mu_e)(\mu_{_M}+\alpha_{_M}\frac{\Lambda}{\mu_e})}$
$a_1a_2-a_3>0 $ implies that 
\begin{eqnarray}
&\psi\frac{\left[(\mu_{_H}+\delta_{_H}+\alpha_{_H }\frac{\Lambda}{\mu_e})(\delta_{_M}+\mu_e+\mu_{_M}+\alpha_{_M}\frac{\Lambda}{\mu_e})+(\delta_{_M}+\mu_e)(\mu_{_M}+\alpha_{_M}\frac{\Lambda}{\mu_e})\right]}{(\mu_{_H}+\delta_{_H}+\alpha_{_H }\frac{\Lambda}{\mu_e})(\delta_{_M}+\mu_e)(\mu_{_M}+\alpha_{_M}\frac{\Lambda}{\mu_e})}(1-R_{_1}^2)\\
&>(1-R^2_{_{2}}).\end{eqnarray} 
This is greater than zero $(> 0)$ if and only if $R_{_1}^2<1$ and $R_{_2}^2 <1,$ where $\psi=(\delta_{_M}+\mu_e+\mu_{_M}+\alpha_{_M}\frac{\Lambda}{\mu_e}+\mu_{_H}+\delta_{_H}+\alpha_{_H }\frac{\Lambda}{\mu_e})$.
Therefore, all eigenvalues of the system have negative real parts. Hence, the following result is obtained: \begin{lem}The invasion-free state is locally asymptotically stable as long as $R^2_{_{2}} < 1$ and $R_{_1}^2 < 1$.\end{lem}For global stability of the initial invasion state, consider the following Lyapunov function, 
\begin{eqnarray*}
\cal{L}&=&I_e+M+H
\end{eqnarray*} which is positive definite $\forall$ $I_e,~M,~H$ as $t\rightarrow \infty$, with orbital derivative given by
\begin{eqnarray*}\begin{array}{ll}
\cal{L'}=&I_e'+M'+H'\\
=&\alpha_{_M}U_eM+\alpha_{_H }U_eH-\mu_e I_e-\delta_{_M}I_e\\&+r\delta_{_M}(1-\epsilon)I_e-\mu_{_M} M-\alpha_{_M}U_eM+\Pi-\mu_{_H}H-\delta_{_H}H-\alpha_{_H }U_eH\\=&-\mu_e I_e-\delta_{_M}I_e+r\delta_{_M}(1-\epsilon)I_e-\mu_{_M} M M+\Pi-\mu_{_H}H-\delta_{_H}H \\\leq&(r\delta_{_M}(1-\epsilon)-\mu_e-\delta_{_M})I_e-\mu_{_M} M-\mu_{_H}H-\delta_{_H}H\\=&(\mu_e+\delta_{_M})(\frac{r\delta_{_M}(1-\epsilon)}{\mu_e+\delta_{_M}}-1)I_e-\mu_{_M} M-\mu_{_H}H-\delta_{_H}H\\=&(\mu_e+\delta_{_M})(R^2_0-1)I_e-\mu_{_M} M-(\mu_{_H}+\delta_{_H})H,
~\mbox{where $R^2_0=\frac{r\delta_{_M}(1-\epsilon)}{\mu_e+\delta_{_M}}$}\\\leq&~~0 ~~\mbox{if}~~R_{0}^2\leq~1
\end{array}\end{eqnarray*}
It is important to note that helminths larvae invade the blood stream a year after an individual is bitten by the fly. It is therefore biologically meaningful to assume that $H\rightarrow 0$ as $t\rightarrow \infty$. With this assumption, our system has a maximum invariant set for $\cal{L}'$=0 if and only if $R_{0}^2 \leq 1$ holds and there is no new invasion of the red blood cells. Therefore, all the trajectories starting in the feasible region where the solutions have biological meaning approach the positively invariant subset of the set where $\cal{L'}$$=0,$ which is the set where $I_e=M=0$ and $H\rightarrow 0$. In this set $N_e\rightarrow\frac{\Lambda}{\mu_e}$ as $t\rightarrow+\infty$. This shows that all solutions approach the initial invasion steady state. Thus, when $R_{0}^2~\leq~1$, malaria and helminthiasis will be eliminated from the blood circulation. If $R_{0}^2~>~1$, then $\cal{L'}$$>0$ for $I_e,~M ,N_{e},H$ close to $(0,0,\frac{\Lambda}{\mu_e},0)$  in $\Gamma$ except $I_e=0$. Thus, the following conclusion is made: \begin{lem}The initial invasion state is globally asymptotically stable in $\Gamma$ if $R_{0}^2~\leq~1$. In this case, the co-infection of malaria and helminthiasis will be kept under control. This would mean that fewer merozoites will be released per rupturing infected erythrocyte.\label{lem3}\end{lem}When $R_{0}^2~>~1~\mbox{then}~\frac{\delta_{_M}[r(1-\epsilon)]}{(\mu_e+\delta_{_M})} >1 ~\mbox{implying that}~~ (1-\epsilon) > \frac{(\mu_e+\delta_{_M})}{r\delta_{_M}}$. Thus, $\epsilon <  1-\frac{(\mu_e+\delta_{_M})}{r\delta_{_M}}$. This gives a percentage value of drug action that is sufficient for treatment of the co-infection of malaria and visceral helminthiasis in absence of immune response. Thus we have the following result: \begin{lem}$\epsilon_o =  1-\frac{(\mu_e+\delta_{_M})}{r\delta_{_M}}$ is the threshold value of the drug action necessary and sufficient to treat the co-infection of malaria and visceral helminthiasis in absence of immune response.\label{lem4}\end{lem}
The biological feasible region for system (\ref{helmM2}) is the invariant simplex in the positive cone of $R_+^4$ given by $\Gamma=\{(U_e,I_e,M,H)\in R_+^4 \}$ including all of its lower dimensional boundaries. Mathematically, system (\ref{helmM2}) will be regarded as a system in $R_+^4$ with an invariant manifold $\Gamma$ of dimension $3$. Re-writing (\ref{helmM2}) in the form \begin{eqnarray}\begin{array}{l}\dot{x}=f(x),~~~\dot{z}=\frac{\partial f^{[3]}}{\partial x}(x)z,\end{array}{\label{X}}\end{eqnarray} where $z=(z_1,z_2,z_3,z_4)\in R^4 \cong R^{^{^{\left(\begin{array}{l}4\\3\end{array}\right)}}}$. The third additive compound $\frac{\partial f^{[3]}}{\partial x}$ for system (\ref{helmM2}) using (\ref{J}) is given by \begin{eqnarray}\frac{\partial f^{[3]}}{\partial x}=-[\mu_e+\alpha_{_M}M+\alpha_{_H}H+\delta_{_M}+\mu_e+\mu_{_M}+(\alpha_{_M}+\alpha_{_H})U_e]I+\phi{\label{V1}}\end{eqnarray} where $\phi$ is a matrix given by \begin{eqnarray}\nonumber\phi=\left[
\begin{array}{cccc}\alpha_{_H}U_e&0&-\alpha_{_H}U_e&-\alpha_{_H}U_e
\\0&-\mu_{_H}H-\delta_{_H}H+\alpha_{_M}U_e+\mu_{_M}&\alpha_{_M}U_e&\alpha_{_M}U_e\\
\alpha_{_H}H&-\alpha_{_M}M+r\delta_{_M}(1-\epsilon)&\delta_{_M}+\mu_e-\mu_{_H}H-
\delta_{_H}H&0\\-\alpha_{_H}H&\alpha_{_M}M&\alpha_{_M}M+\alpha_{_H}H&\rho\end{array}\right]\\
{\label{X1}}\end{eqnarray} with $\rho=-\mu_{_H}H-\delta_{_H}H+\mu_e+\alpha_{_M}M+\alpha_{_H}H$. To show asymptotic stability of the endemic equilibrium, we use the method of first approximation \cite{Li}. Using the spectral properties of the second compound matrices, we state the following Lemma. \begin{lem}Let A be an n $\times$ n matrix with real entries. For A to be stable, it is necessary and sufficient that \begin{itemize}\item[1.] the third compound matrix $A^{[3]} $ is stable, \item[2.] $(-1)^ndet(A) >0$.\end{itemize} \end{lem}
\begin{prop} The endemic equilibrium is stable if the following inequalities are satisfied
\begin{eqnarray}\begin{array}{rl}
\Lambda r\delta_{_M}(1-\epsilon)I_e^* >& 2(\alpha_{_M}M^*U_e^*)^2\\\frac{I_e^*}{U_e^*}>& 1\\\delta_{_M}\Pi I_e^*>&2(\alpha_{_H}U_e^*H^*)^2 
\end{array}\end{eqnarray}
that is, if the mean infection burden is greater than 1. 
\end{prop} {\bf{Proof:}}
The third additive compound matrix of System (\ref{helmM2}), $J^{[3]}$, is given by (\ref{V1}). For $E^*=(U_e^*,I_e^*,M^*,H^*)$, and the diagonal matrix $D=\mbox{diag}(U_e^*,I_e^*,M^*,H^*)$, the matrix $J^{[3]}(E^*)$ is similar to $DJ^{[3]}(E^*)D^{-1}$. This is given by \begin{eqnarray}-[\mu_e+\alpha_{_M}M^*+\alpha_{_H}H^*+\delta_{_M}+\mu_e+\mu_{_M}+(\alpha_{_M}+\alpha_{_H})U_e^*]I+\psi{\label{V2}}
\end{eqnarray} where $\psi$ is the following matrix
\begin{eqnarray}\nonumber\psi=\left[\begin{array}{cccc}\alpha_{_H}U_e^*&0&\frac{-\alpha_{_H}U_e^*M^*}{U_e^*}&
\frac{-\alpha_{_H}U_e^*H^*}{U_e^*}\\0&-\mu_{_H}H-\delta_{_H}H+\alpha_{_M}U_e^*+\mu_{_M}&\frac{\alpha_{_M}U_e^*M^*}{I_e^*}&\frac{\alpha_{_M}U_e^*H^*}{I_e^*}\\\frac{\alpha_{_H}H^*U_e^*}{M^*}&\frac{(-\alpha_{_M}M^*+r\delta_{_M}(1-\epsilon))I_e^*}{M^*}&\delta_{_M}+\mu_e-\mu_{_H}H-\delta_{_H}H&0\\\frac{-\alpha_{_H}H^*U_e}{H^*}&\frac
{\alpha_{_M}M^*I_e^*}{H^*}&\frac{(\alpha_{_M}M^*+\alpha_{_H}H^*)M^*}{H^*}&\theta\end{array}\right]{\label{X2}}\\
\end{eqnarray} with $\theta=-\mu_{_H}H-\delta_{_H}H+\mu_e+\alpha_{_M}M^*+\alpha_{_H}H^*$. The matrix $J^{[3]}(E^*)$ is stable if and only if $DJ^{[3]}(E^*)D^{-1}$ is stable, for similarity preserves the eigenvalues. Since the diagonal elements of the matrix $DJ^{[3]}(E^*)D^{-1}$ are negative, an easy argument using Ger$\bar{s}$gorin discs shows that it is stable if it is diagonally dominant in rows, \cite{Us}. Set $\mu=\{g_1,g_2,g_3,g_4\}$ where \begin{eqnarray}\begin{array}{ll}
g_1=&-\left[\mu_e+\alpha_{_M}M^*+\alpha_{_H}H^*+\delta_{_M}+\mu_e+\mu_{_M}+\alpha_{_M}U_e^*+\alpha_{_H}(M^*+H^*)\right]\\
g_2=&-[\mu_e+\alpha_{_M}M^*+\alpha_{_H}H^*+\delta_{_M}+\mu_e+\alpha_{_H}U_e^*]-\mu_{_H}H-\delta_{_H}H+\frac{\alpha_{_M}U_e^*M^*}{I_e^*}+\frac{\alpha_{_M}U_e^*H^*}{I_e^*}\\
g_3=&-[\mu_e+\alpha_{_M}M^*+\alpha_{_H}H^*+\mu_{_M}+(\alpha_{_M}+\alpha_{_H})U_e^*]\\&+\frac{\alpha_{_H}H^*U_e^*}{M^*}+\frac{(-\alpha_{_M}M^*+r\delta_{_M}(1-\epsilon))I_e^*}{M^*}-\mu_{_H}H-\delta_{_H}H\\
g_4=&-[\delta_{_M}+\mu_e+\mu_{_M}+(\alpha_{_M}+\alpha_{_H})U_e^*]\\&-\frac{\alpha_{_H}H^*U_e}{H^*}+\frac{\alpha_{_M}M^*I_e^*}{H^*}+\frac{(\alpha_{_M}M^*+\alpha_{_H}H^*)M^*}{H^*}-\mu_{_H}H-\delta_{_H}H
\end{array}{\label{G5}}\end{eqnarray}
At steady state, System (\ref{helmM2}) can be written as 
\begin{eqnarray}\begin{array}{ll}\frac{1}{2}[\frac{\Lambda}{U_e^*}-\mu_e+(\mu_e +\delta_{_M})\frac{I_e^*}{U_e^*}]&=(\alpha_{_M}M^*+\alpha_{_H }H^*)\\r\delta_{_M}(1-\epsilon)\frac{I_e^*}{M^*}-\mu_{_M} &=\alpha_{_M}U_e^*\\\frac{\Pi}{H^*}-\mu_{_H}H-\delta_{_H}H &=\alpha_{_H }U_e^*\\\end{array}{\label{Y}}\end{eqnarray}
Therefore, substituting Equation (\ref{Y}) in (\ref{G5}) we obtain 
\begin{eqnarray}\begin{array}{ll}
g_1=&-\left[\frac{\mu_e}{2}+\frac{\Lambda}{2U_e^*}+\mu_e(1+\frac{I_e^*}{2U_e^*})+\delta_{_M}[1+I_e^*(\frac{1}{2U_e^*}+\frac{r}{M^*})]+\alpha_{_H}(M^*+H^*)\right]\\
g_2=&-\left[\frac{\mu_e}{2}+\frac{\Lambda}{2U_e^*}+\mu_e(1+\frac{I_e^*}{2U_e^*})+\delta_{_M}(1+\frac{I_e^*}{2U_e^*})-\frac{\alpha_{_M}U_e^*}{I_e^*}(M^*+H^*)\right]\\
g_3=&-\left[\frac{\mu_e}{2}+\frac{\Lambda}{2U_e^*}+\frac{1}{2}(\mu_e+\delta_{_M})\frac{I_e^*}{U_e^*}+\frac{\Pi}{H^*}+\alpha_{_M}I_e^*-\frac{\alpha_{_H}U_e^*H^*}{M}\right]\\
g_4=&-\left[\delta_{_M}[1+\frac{rI_e^*}{M^*}-\frac{I_e^*M^*}{2H^*U_e^*}]+\mu_e(1-\frac{M^*I_e^*}{2H^*U_e^*})+\frac{\Pi}{H^*}+\alpha_{_H}U_e^*+\frac{M^*}{2H^*}[\mu_e-\frac{\Lambda }{U_e^*}]-\frac{\alpha_{_M}M^*I_e^*}{H^*}\right]
\end{array}\end{eqnarray}
This gives $\mu <0$ which implies diagonal dominance as claimed and thus verifies the first condition.\\

 From Jacobian \ref{J} and Equation (\ref{Y}) , 
\begin{footnotesize}\begin{eqnarray}det[J(E^*)]&=&\nonumber\left|\begin{array}{cccc}-\mu_e-\alpha_{_M}M^*-\alpha_{_H }H^*&0&-\alpha_{_M}U_e^*&-\alpha_{_H }U_e^*\\\alpha_{_M}M^*+\alpha_{_H }H^*&-\delta_{_M}-\mu_e&\alpha_{_M}U_e^*&\alpha_{_H }U_e^*\\-\alpha_{_M}M^*&-\alpha_{_M}M^*+r\delta_{_M}(1-\epsilon)&-\mu_{_M}-\alpha_{_M}U_e^*&0\\-\alpha_{_H }H^*&-\alpha_{_H }H^*&0&-\mu_{_H}-\alpha_{_H }U_e^*\end{array}\right|\\&=&\nonumber\left|\begin{array}{cccc}-\frac{1}{2}[\frac{\Lambda}{U_e^*}-\mu_e+(\mu_e +\delta_{_M})\frac{I_e^*}{U_e^*}]&0&-\alpha_{_M}U_e^*&-\alpha_{_H }U_e^*\\\frac{1}{2}[\frac{\Lambda}{U_e^*}-\mu_e+(\mu_e +\delta_{_M})\frac{I_e^*}{U_e^*}]&-\delta_{_M}-\mu_e&\alpha_{_M}U_e^*&\alpha_{_H }U_e^*\\-\alpha_{_M}M^*&-\alpha_{_M}M^*+r\delta_{_M}(1-\epsilon)&-\frac{r\delta_{_M}(1-\epsilon)I_e^*}{M^*}&0\\-\alpha_{_H }H^*&-\alpha_{_H }H^*&0&-\frac{\Pi}{H^*}\end{array}\right|\\&=&\nonumber\left\{\frac{\Pi}{H^*}(\delta_{_M}+\mu_e)\left[\frac{1}{2}[\frac{\Lambda}{U_e^*}-\mu_e+(\mu_e +\delta_{_M})\frac{I_e^*}{U_e^*}]\right]\left[\frac{r\delta_{_M}(1-\epsilon)I_e^*}{M^*})\right]\right\}\\&-&\left\{\frac{\Pi}{H^*}(\alpha_{_M}^2M^*U_e^*)(\delta_{_M}+\mu_e)+(\alpha_{_H }H^*)(\delta_{_M}+\mu_e)(\alpha_{_H }U_e^*)(\frac{r\delta_{_M}(1-\epsilon)I_e^*}{M^*})\right\}\end{eqnarray}\end{footnotesize}\begin{eqnarray}\begin{array}{ll}=&(\delta_{_M}+\mu_e)\{\frac{\Pi\Lambda r\delta_{_M}(1-\epsilon)I_e^*}{2H^*U_e^*M^*}+\frac{r\delta_{_M}(1-\epsilon)\mu_e\Pi I_e^{*2}}{2H^*M^*U_e^*}+\frac{r\delta_{_M}(1-\epsilon)^2\Pi I_e^{*2}}{2H^*M^*U_e^*}\\&-\frac{r\delta_{_M}(1-\epsilon)\mu_e\Pi I_e^*}{2H^*M^*}-\frac{\Pi\alpha^2_{_M}M^*U_e^*}{H^*}-\frac{r\delta_{_M}(1-\epsilon)\alpha_{_H}^2U_e^*I_e^*H^*}{M^*}\}
\\=&(\delta_{_M}+\mu_e)\{\frac{\Pi\alpha_{_M}^2M^*U_e^*}{H^*}\left[\frac{\Lambda r\delta_{_M}(1-\epsilon)I_e^*}{2\alpha_{_M}^2M^{*2}U_e^{*2}}-1\right]+\frac{r\mu_e\delta_{_M}\Pi I_e^* }{2H^*M^*}\left[\frac{\mu_eI_e^*}{\mu_eU_e^*}-1\right]\\&+\frac{r\delta_{_M}(1-\epsilon)\alpha_{_H}^2I_e^*U_e^*H^*}{M^*}\left[\frac{\delta_{_M}\Pi I_e^*}{2\alpha_{_H}^2U_e^{*2}H^{*2}}-1\right]\}
\end{array}\end{eqnarray}
Note that in the first expression, $\frac{r\delta_{_M}(1-\epsilon)I_e^*}{\alpha_{_M}MU_e}$ is the ratio of the net rate of merozoite production $r\delta_{_M}(1-\epsilon) I_e$ per ruptured infected cell in presence of drugs to the total number of free merozoites that invade susceptible erythrocytes $\alpha_{_M} U_e M$, while $\frac{\Lambda}{\alpha_{_M}MU_e}$ is the invasion rate per recruited erythrocyte. The second expression $\frac{I_e^*}{U_e^*}$ gives the mean infection burden of erythrocytes, while the third expression $\frac{\Pi\delta_{_M} I_e^*}{2\alpha_{_H}^2U_e^{*2}H^{*2}}$, gives the net rate of helminths invasion of red blood cells per released larvae $\frac{\Pi }{\alpha_{_H}U_e^*H^*}$ and the net rate of rupture of infected erythrocytes in the presence of helminths invasion $\frac{\delta_{_M}I_e^*}{\alpha_{_H}U_e^*H^*}$ at endemic equilibrium. Therefore, $det[J(E^*)] > 0$ and this completes the proof. $\Box$

\subsection{Criteria for invasion and persistence in the system with immune response} 
\begin{figure}[ht!]\centering{
        \includegraphics[width=0.40\textwidth]{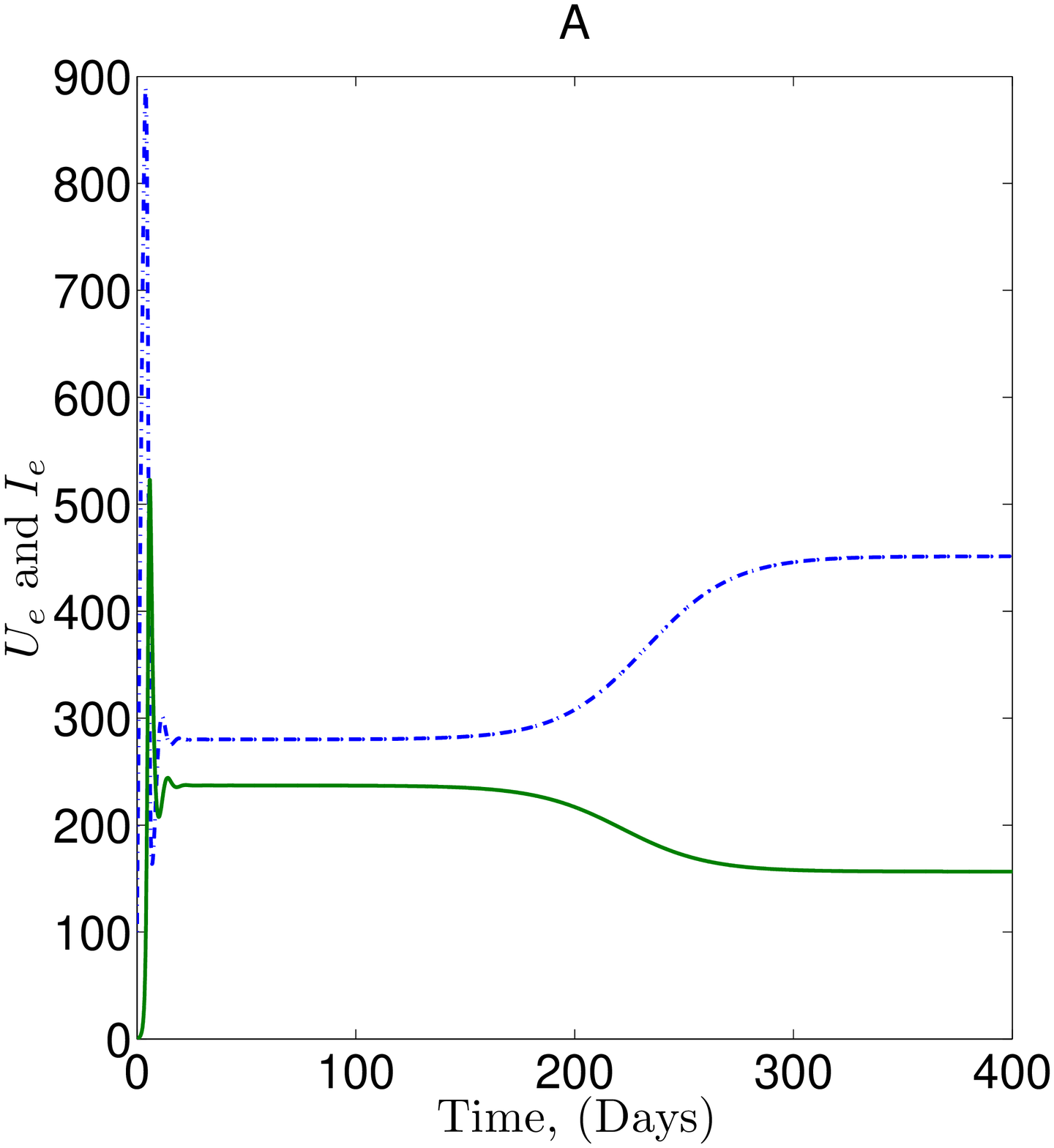} 
       \includegraphics[width=0.40\textwidth]{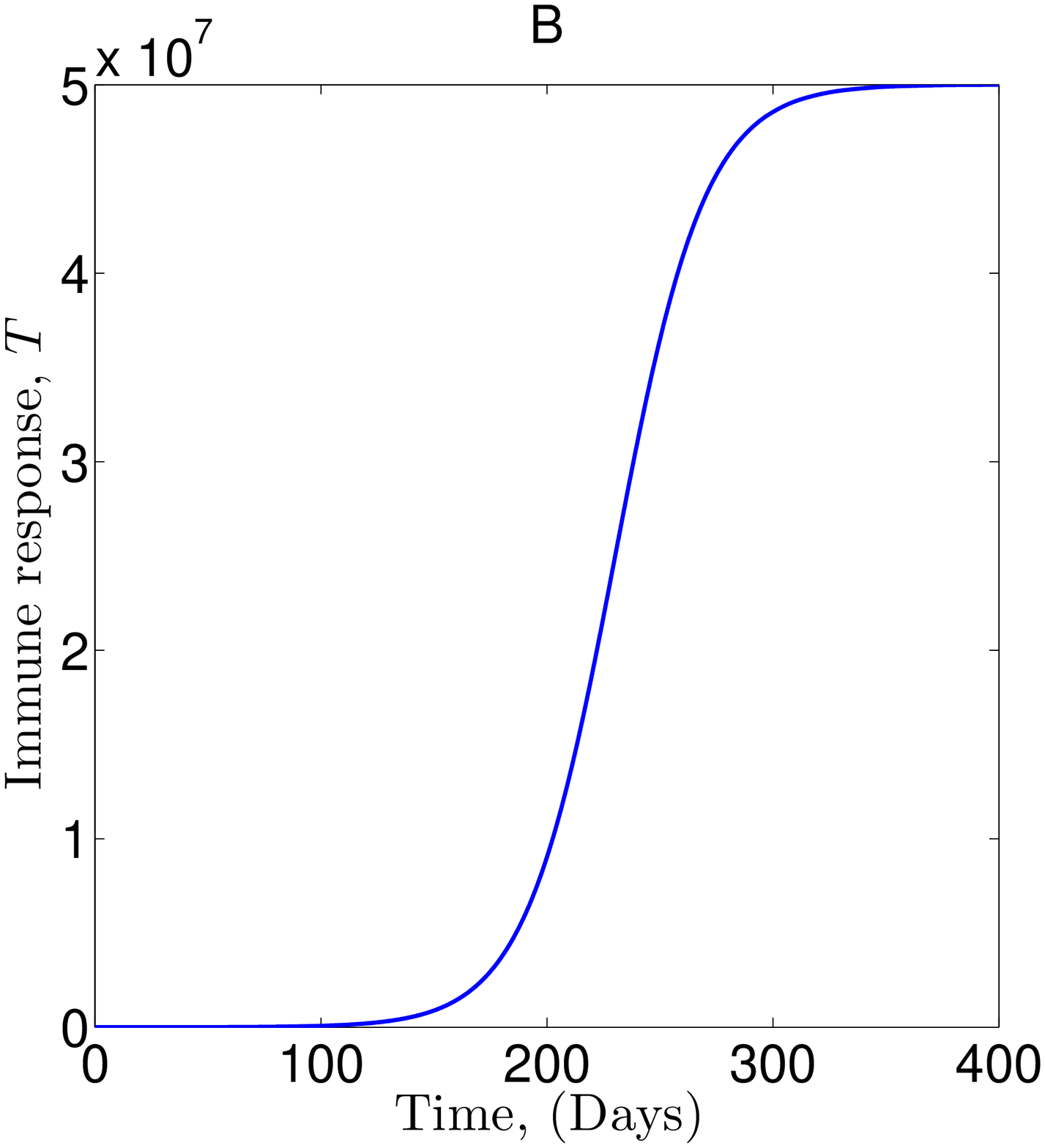}  
       \includegraphics[width=0.40\textwidth]{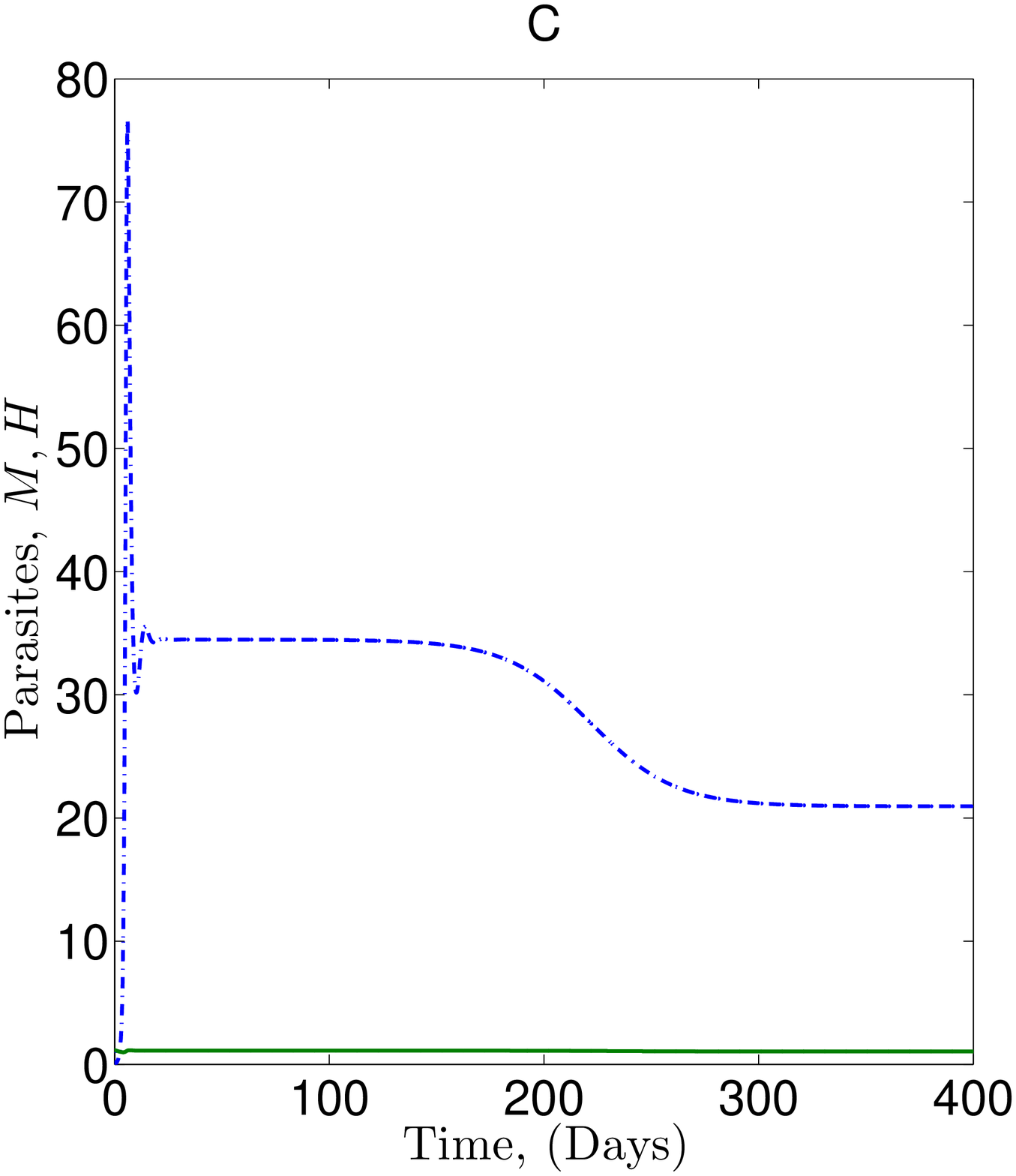}  }
\caption{Graphs for the evolution of uninfected and infected erythrocytes with time (A); the dashed line is for un infected erythrocytes, while the solid line represents the infected erythrocytes. In (B), the evolution of immune response with time is shown, while in (C), evolution of malaria $M$ and helminth larvae $H$ with time is given. In the graph, the dashed line malaria parasites while the solid line represents the helminths larva. Parameter values used are
$\epsilon=0.5, \Lambda=250, \mu_e0.025, \mu_h=20, \mu_M=48, \delta_M=1, \delta_H=1, \alpha_M=0.025, \alpha_H=0.005, r=16, \Pi=25, \gamma_M=0.1, \gamma_I=0.1, \gamma_H=0.1, p=0.1, b=0.000000001, \mu_t=0.05, \lambda_I=0.00000001, \lambda_M=0.00000001, \lambda_H=0.00000001. $\label{helfig2}}\end{figure}

Consider the case when there is immune response to the infections. In this case, we have three categories of equilibrium: the first is the initial invasion state when the number of helminths larvae released in the blood is small and hence the host is not displaying any symptoms and the immune response is not triggered. This is the naive equilibrium where $U_e=\frac{\Lambda}{\mu_e}$ and $I_e = M = T = 0$, $H \approx 0$. Secondly when $\frac{dT}{dt}=0,$ there is a second equilibrium point where $U_e=\frac{\Lambda}{\mu_e}$ and $I_e = M = 0,$ and $H\approx 0 $ but $T = \frac{p-\mu_t}{b} $. At this steady state, the immune system has been triggered and the malaria parasites and helminths larvae have been eliminated and the system has returned to an infection-free state with a degree of residual immunity. The third type of equilibrium represents host-parasite co-existence and is characterized by non-zero levels of $I_e$, $M$ and $H$, $U_e <\frac{\Lambda}{\mu_e}$ and \begin{eqnarray}T>\frac{p-\mu_t}{b}.{\label{IM}}\end{eqnarray} 
The endemic state is too complicated to be useful. However, numerical integration shown in Figure (\ref{helfig2}) give an example of how this endemic state can be attained. We make the following conclusion: \begin{lem}In the presence of immune response, there are three equilibria states.  The first is the initial invasion state where the helminths larvae die off without being ingested by a feeding fly. The second steady state is when the immune population has been triggered and the malaria parasites and helminths larvae have been eliminated. In this case the system returns to an infection-free state with a degree of residual immunity. The third state represents non zero values of $U_e^*,~I_e^*,~M^*$ and $H^*$. \label{lem6}\end{lem} From this result, we have: \begin{lem} In the absence of infection, the immune population (and therefore the naive equilibrium) is inherently unstable and will not return to zero once it has been triggered by the presence of malaria parasites and helminths larvae. \label{lem7}\end{lem}

From Figure \ref{helfig2}, we note a delayed immune response and thus an increase in pathogen concentration. Once the immune response is triggered, there is a significant reduction in $I_e,~M$ and $H$. \\

Next we analyze stability of the states by introducing small perturbations around these states. The Jacobian of the System (\ref{helmal1}) is given by \begin{eqnarray}J_e=\left[\begin{array}{cc}J_{11}&J_{12}\\J_{21}&J_{22}\end{array}\right]{\label{J1}}\end{eqnarray}
where \begin{footnotesize}\begin{eqnarray*}J_{11}=\left[\begin{array}{cc}-\mu_e-\alpha_{_M}M^*-\alpha_{_H }H^*&0\\\alpha_{_M}M^*+\alpha_{_H }H^*&-\delta_{_M}-\mu_e-\lambda_IT^*\end{array}\right],~J_{12}=\left[\begin{array}{ccc}-\alpha_{_M}U_e^*&-\alpha_{_H }U_e^*&0\\\alpha_{_M}U_e^*&\alpha_{_H }U_e^*&-\lambda_II_e^*\\-\mu_{_M}-\alpha_{_M}U_e^*-\lambda_MT^*&0&-\lambda_MM^*\end{array}\right]\end{eqnarray*}
\begin{eqnarray*}J_{21}=\left[\begin{array}{cc}-\alpha_{_M}M^*&-\alpha_{_M}M^*+r\delta_{_M}(1-\epsilon)\\-\alpha_{_H }H^*&-\alpha_{_H }H^*\\0&\gamma_{_I}\end{array}\right],~J_{22}=\left[\begin{array}{ccc}0&-\mu_{_H}-\delta_{_H}-\alpha_{_H}U_e^*-\lambda_{H}T^*&-\lambda_{H}H^*\\\gamma_{_M}&\gamma_{_H}&p-\mu_t-2bT^*\end{array}\right]\end{eqnarray*}\end{footnotesize}
If when $U_e^o=\frac{\Lambda}{\mu_e}$,  $T^*=\frac{p-\mu_t}{b}$, we have
\begin{footnotesize}\begin{eqnarray}\nonumber J_{o}=\left[\begin{array}{ccccc}-\mu_e&0&-\alpha_{_M}\frac{\Lambda}{\mu_e}&-\alpha_{_H }\frac{\Lambda}{\mu_e}&0\\0&-\delta_{_M}-\mu_e-\lambda_I\frac{p-\mu_t}{b}&\alpha_{_M}\frac{\Lambda}{\mu_e}&\alpha_{_H }\frac{\Lambda}{\mu_e}&0\\0&r\delta_{_M}(1-\epsilon)&-\mu_{_M}-\alpha_{_M}\frac{\Lambda}{\mu_e}-\lambda_M\frac{p-\mu_t}{b}&0&0\\0&0&0&-\mu_{_H}-\alpha_{_H }\frac{\Lambda}{\mu_e}-\lambda_H\frac{p-\mu_t}{b}&0\\0&\gamma_{_I}&\gamma_{_M}&\gamma_{_H}&-(p-\mu_t)\end{array}\right]\\\end{eqnarray} 
\end{footnotesize}with distinct eigenvalues $-\mu_e$ and $-(p-\mu_t)$ since $p \gg \mu_t$. The remaining three are given by a 3$\times$3 square matrix given by 
\begin{eqnarray}\nonumber\left[\begin{array}{ccc}-\delta_{_M}-\mu_e-\lambda_I\frac{p-\mu_t}{b}&\alpha_{_M}\frac{\Lambda}{\mu_e}&\alpha_{_H }\frac{\Lambda}{\mu_e}\\r\delta_{_M}(1-\epsilon)&-\mu_{_M}-\alpha_{_M}\frac{\Lambda}{\mu_e}-\lambda_M\frac{p-\mu_t}{b}&0\\0&0&-\mu_{_H}-\alpha_{_H }\frac{\Lambda}{\mu_e}-\lambda_H\frac{p-\mu_t}{b}\end{array}\right]\\\end{eqnarray} with negative eigenvalues $-[\mu_{_H}+\alpha_{_H }\frac{\Lambda}{\mu_e}+\lambda_H\frac{p-\mu_t}{b}]$ and $-\frac{1}{2}\left[a+e+\sqrt{a^2+e^2-2ae+4df}\right]$ where $a=(\delta_{_M}+\mu_e+\lambda_I\frac{p-\mu_t}{b}),$ $e=(\mu_{_M}+\alpha_{_M}\frac{\Lambda}{\mu_e}+\lambda_M\frac{p-\mu_t}{b})$, $d=\alpha_{_M}\frac{\Lambda}{\mu_e}$ and $f=r\delta_{_M}(1-\epsilon)$ respectively. The last eigenvalue given by $-\frac{1}{2}\left[a+e-\sqrt{a^2+e^2-2ae+4df}\right]$ has negative real part if and only if \begin{eqnarray*}\begin{array}{rl}a+e &> \sqrt{a^2+e^2-2ae+4df}\\
\mbox{that is,} ~~(\delta_{_M}+\mu_e+\lambda_I\frac{p-\mu_t}{b})(\mu_{_M}+\alpha_{_M}\frac{\Lambda}{\mu_e}+\lambda_M\frac{p-\mu_t}{b}) &> (\alpha_{_M}\frac{\Lambda}{\mu_e})(r\delta_{_M}(1-\epsilon))\\
1-\frac{(\alpha_{_M}\frac{\Lambda}{\mu_e})(r\delta_{_M}(1-\epsilon))}{(\delta_{_M}+\mu_e+\lambda_I\frac{p-\mu_t}{b})(\mu_{_M}+\alpha_{_M}\frac{\Lambda}{\mu_e}+\lambda_M\frac{p-\mu_t}{b})}&>0\\
1-R^2_{2_i}&>0~~ \mbox{if $R^2_{2_i} < 1$}\end{array}
\end{eqnarray*} where $R^2_{2_i}=\frac{\alpha_{_M}r\delta_{_M}\frac{\Lambda}{\mu_e}(1-\epsilon)}{(\delta_{_M}+\mu_e+\lambda_I\frac{p-\mu_t}{b})(\mu_{_M}+\alpha_{_M}\frac{\Lambda}{\mu_e}+\lambda_M\frac{p-\mu_t}{b})}.$ The term $\frac{\alpha_{_M}\frac{\Lambda}{\mu_e}}{(\delta_{_M}+\mu_e+\lambda_I\frac{p-\mu_t}{b})}$ gives the ratio of the infection to the total mortality rate of the erythrocytes while $\frac{r\delta_{_M}(1-\epsilon)}{(\mu_{_M}+\alpha_{_M}\frac{\Lambda}{\mu_e}+\lambda_M\frac{p-\mu_t}{b})}$ is the rate of release of merozoites against drug and immune action per dying merozoite . This means that $R^2_{2_i}$ is the within-host reproductive potential of a merozoite released in presence of helminths larvae, against drug and immune action. Thus, the following conclusion is made:  \begin{lem}The initial invasion state, the immune equilibrium where $T=\frac{p-\mu_t}{b}$ is stable if \begin{eqnarray}\frac{\alpha_{_M}r\delta_{_M}\frac{\Lambda}{\mu_e}(1-\epsilon)}{(\delta_{_M}+\mu_e+\lambda_I\frac{p-\mu_t}{b})(\mu_{_M}+\alpha_{_M}\frac{\Lambda}{\mu_e}+\lambda_M\frac{p-\mu_t}{b})} < 1.{\label{MY}}\end{eqnarray}\end{lem} This biologically important criterion determines how malaria parasites or helminths larvae following a successful invasion, will eventually be controlled by the immune response. If the criterion is not satisfied, neither the malaria parasites nor the helminths larvae will be eliminated but either or both will persist at an equilibrium density within the host. Alternatively, if a host undergoes successful therapy of either infection or both, and the immune cells have settled at their equilibrium level, this criterion determines whether a new infection is able to overcome the stability of this equilibrium and re-infect the host. The outcome of infection depends upon the balance between malaria parasite and helminths larvae and host attributes. As the death rate of the merozoites increases, the immune equilibrium becomes more and more stable and the host is able to eliminate the infection more easily. Similarly, the higher the equilibrium level of immunity $\frac{p-\mu_t}{b}$, the greater the malaria parasite invasion rate $\alpha_{_M}$ must be to achieve reinfection. It is interesting to note that in this system, the stability of the immune equilibrium does not depend upon the rates of $T$ cell activation by the pathogens. However, the speed at which the system moves towards equilibrium will be influenced by the magnitude of these rates. \\

For global stability of the initial state equilibrium with immunity, consider the following Lyapunov function, 
\begin{eqnarray*}
\cal{L}&=&I_e+M+H+T
\end{eqnarray*} which is positive definite $\forall$ $I_e,~M,~H,T$ as $t\rightarrow \infty$, with orbital derivative given by
\begin{eqnarray*}\begin{array}{ll}
\cal{L'}=&I_e'+M'+H'+T'\\
=&\alpha_{_M}U_eM+\alpha_{_H }U_eH-\mu_e I_e-\delta_{_M}I_e-\lambda_{_I}I_e T +r\delta_{_M}(1-\epsilon)I_e-\mu_{_M} M\\&-\alpha_{_M}U_eM-\lambda_{_M} M T+\Pi-\mu_{_H}H-\delta_{_H}H-\alpha_{_H }U_e-\lambda_{_H} H T+\gamma_{_M} M\\&+\gamma_{_I} I_e+\gamma_{_H} H+p T-\mu_t T-bT^2\\=&-\mu_e I_e-\delta_{_M}I_e+r\delta_{_M}(1-\epsilon)I_e-\mu_{_M} M+\Pi-\mu_{_H}H-\delta_{_H}H-\alpha_{_H }I_e H\\&+(\gamma_{_M}-\lambda_{_M} T)M+(\gamma_{_H}-\lambda_{_H} T)H+\gamma_{_I} I_e-\lambda_{_I}I_eT+p T-\mu_t T-bT^2\\\leq&r\delta_{_M}(1-\epsilon)+\gamma_{_I}-(\mu_e+\delta_{_M}+\lambda_{_I}T)I_e+(p-\mu_t-bT)T\\=&(\mu_e
+\delta_{_M}+\lambda_{_I}T)(\frac{r\delta_{_M}(1-\epsilon)+\gamma_{_I}}{(\mu_e+\delta_{_M}+\lambda_{_I}T)}-1)I_e
+bT^2(\frac{p-\mu_t}{bT}-1)\\=&(\mu_e+\delta_{_M}+\lambda_{_I}T)(R^2_i-1)I_e+bT^2(\frac{p-\mu_t}{bT}-1)~\mbox{where $R^2_i=\frac{r\delta_{_M}(1-\epsilon)+\gamma_{_I}}{(\mu_e+\delta_{_M}+\lambda_{_I}T)}$}\\\leq&~~0 ~~\mbox{if}~~R_{i}^2\leq~1~~\mbox{and since $\frac{p-\mu_t}{bT} \leq 1$} 
\end{array}\end{eqnarray*} $R_i$	 is defined as the within-host rupture potential of an infected erythrocyte with immune action. Thus, the following result is obtained: \begin{lem}In presence of immune response, the initial invasion state is globally asymptotically stable in $\Gamma$ if $R_{i}^2~\leq~1$. In this case, the immune population has been triggered but the malaria parasites and helminths larvae have been eliminated and the system has returned to an infection-free state with a degree of residual immunity.\end{lem}This further leads to the following conclusion:  \begin{lem}It is further necessary that to avoid reinfection from malaria, $\frac{p-\mu_t}{b} \ll 1$ since the higher the equilibrium level of immunity $\frac{p-\mu_t}{b}$, the greater the malaria parasite invasion rate $\alpha_{_M}$ in presence of helminths larvae must be to achieve reinfection.\end{lem} When $R_{i}^2~>~1$, it implies that 
\begin{eqnarray*}\begin{array}{rl}\frac{r\delta_{_M}(1-\epsilon)+\gamma_{_I})}{(\mu_e+\delta_{_M}+\lambda_{_I}T)}>&1\\
1-\epsilon>&\frac{1}{r\delta_{_M}}(\mu_e+\delta_{_M}+\lambda_{_I}T)-\gamma_{_I}\\
\epsilon<&1-\frac{1}{r\delta_{_M}}(\mu_e+\delta_{_M}+\lambda_{_I}T)+\gamma_{_I}\end{array}\end{eqnarray*} 
This expression gives the critical level of drug action below which no recovery is expected. 
Thus we have the following result:\begin{thm}The percentage threshold value of the drug action that is sufficient for recovery in presence of immune response, $\epsilon_i$ is given by \begin{eqnarray}\epsilon_i=1-\frac{1}{r\delta_{_M}}(\mu_e+\delta_{_M}+\lambda_{_I}T)+\gamma_{_I}\end{eqnarray}\end{thm} Note that $\epsilon_i < \epsilon_o$. Conversely, the following counter result is achieved: \begin{thm}The maximum value of $T$-cells required to neutralize the antigens to eliminate a co-infection of malaria and helminthiasis in presence of therapy, $T_{max}$ is given by \begin{eqnarray} T_{max}
=\frac{1}{\lambda_{_I}} (r\delta_{_M}(1-\epsilon)+\gamma_{_I}-\mu_e-\delta_{_M})\end{eqnarray}\end{thm}
For endemicity, from (\ref{MY}), when $T=\frac{p-\mu_t}{b}$ we have $cd-(a+\lambda_IT)(b+\lambda_MT) > 0$ where $a=\delta_{_M}+\mu_e,~b=\mu_{_M}+\alpha_{_M}\frac{\Lambda}{\mu_e},~c=\alpha_{_M}\frac{\Lambda}{\mu_e}$ and $d=r\delta_{_M}(1-\epsilon).$ This gives $(a+\lambda_IT)(b+\lambda_MT)-cd < 0$ implying that $\lambda_{_I}\lambda_{_M}T^2+(b\lambda_{_I}+a\lambda_{_M})T+(ab-cd)<0$ and hence, $f(T)=AT^2+BT+C<0$. At steady state, the non-zero response of immune response satisfies $f(T)=AT^2+BT+C=0$, and this can be analyzed for the possibility of multiple immune responses. Note that the coefficient $A=\lambda_{_I}\lambda_{_M} >0$ and $B=\lambda_{_I}(\mu_{_M}+\alpha_{_M}\frac{\Lambda}{\mu_e})+\lambda_{_M}(\delta_{_M}+\mu_e) > 0$. Hence the following result is established: \begin{thm}There exists \begin{itemize}\item[(i)] precisely one positive immune response if $C= [(\delta_{_M}+\mu_e)(\mu_{_M}+\alpha_{_M}\frac{\Lambda}{\mu_e})-\alpha_{_M}r\delta_{_M}\frac{\Lambda}{\mu_e}(1-\epsilon)] < 0$, \item[(ii)] precisely one unique immune response if $B^2-4AC=$ \begin{footnotesize}$[\lambda_{_I}(\mu_{_M}+\alpha_{_M}\frac{\Lambda}{\mu_e})+\lambda_{_M}(\delta_{_M}+\mu_e)]^2-4\lambda_{_I}\lambda_{_M}[(\delta_{_M}+\mu_e)(\mu_{_M}+\alpha_{_M}\frac{\Lambda}{\mu_e})-r\delta_{_M}\alpha_{_M}\frac{\Lambda}{\mu_e}(1-\epsilon)]=0$,\end{footnotesize}\item[(iii)]precisely two immune responses if $B^2-4AC>0$,\item[(iv)]The immune response is impaired otherwise.\end{itemize}\end{thm}
The following proposition states the local stability of the endemic equilibrium. \begin{prop}The endemic equilibrium is locally asymptotically stable whenever it exists. \end{prop} 
{\bf{Proof:}}\\
At endemic equilibrium, from Jacobian \ref{J1}, it is noted that since the diagonal entries of the Jacobian $J_e$ are negative and the eigenvalues of any square matrix are the same as those of its transpose, then using Ger$\bar{s}$gorin discs \cite{Li}, it suffices to show that $J_e$ is stable if it is diagonally dominant in columns. Let $\mu=\max \{g_1,g_2,g_3,g_4,g_5\}  $ where \begin{eqnarray}\begin{array}{lll}
g_1&=&-\mu_e-\alpha_{_M}M^*-\alpha_{_H }H^*+\alpha_{_M}M^*+\alpha_{_H }H^*-\alpha_{_M}M^*-\alpha_{_H }H^*\\&=&-\mu_e-\alpha_{_M}M^*-\alpha_{_H }H^*\\&<&0\end{array}\end{eqnarray}\begin{eqnarray}\begin{array}{lll}
g_2&=&-\delta_{_M}-\mu_e-\lambda_IT^*-\alpha_{_M}M^*+r\delta_{_M}(1-\epsilon)-\alpha_{_H }H^*+\gamma_{_I}\\&=&-\alpha_{_M}M^*-\alpha_{_H }H^*+(\mu_e+\delta_{_M}+\lambda_IT^*)(R^2_i-1),\\&<&0\\
g_3&=&-\alpha_{_M}U_e^*+\alpha_{_M}U_e^*-\mu_{_M}-\alpha_{_M}U_e^*-\lambda_MT^*+\gamma_{_M}\\&=&-\mu_{_M}-\alpha_{_M}U_e^*-\lambda_MT^*+\gamma_{_M} \\&<& 0~\mbox{since $\mu_{_M} > \gamma_{_M}$; (refer to Hethcote, 2000)}.\\
g_4&=&-\alpha_{_H }U_e^*+\alpha_{_H }U_e^*-\mu_{_H}-\alpha_{_H }U_e^*-\lambda_HT^*+\gamma_{_H}=\\&=&-\mu_{_H}-\alpha_{_H }(U_e^*+I_e^*)-\lambda_HT^*+\gamma_{_H}\\&<&0~ \mbox{since $\mu_{_H} > \gamma_{_H}$}\\
g_5&=&-\lambda_II_e^*-\lambda_MM^*-\lambda_HH^*+p-\mu_t-2bT\\&=&-\lambda_II_e^*-\lambda_MM^*-\lambda_HH^*-2bT+p-\mu_t
\\&=&-\lambda_II_e^*-\lambda_MM^*-\lambda_HH^*-2bT(1-\frac{p-\mu_t}{2bT})\\&<& 0 ~\mbox{since $T>\frac{p-\mu_t}{2b}$ ~from equation (\ref{IM}).}
\end{array}\end{eqnarray} 
$R_i^2=\frac{r\delta_{_M}(1-\epsilon)+\gamma_{_I}}{\mu_e+\delta_{_M}+\lambda_IT^*}$ is the basic infection induced rupture rate with immune activation per dying erythrocyte, then, $R^2_i < 1$ for stability. This implies that \begin{eqnarray}\nonumber \mu&=&\max \{-\mu_e-\alpha_{_M}M^*-\alpha_{_H }H^*,-\alpha_{_M}M^*-\alpha_{_H }H^*+(\mu_e+\delta_{_M}+\lambda_IT^*)(R^2_i-1),\\&&-\mu_{_M}-\alpha_{_M}U_e^*-\lambda_MT^*+\gamma_{_M},-\mu_{_H}-\alpha_{_H }U_e^*-\lambda_HT^*+\gamma_{_H},\\&&\nonumber-\lambda_II_e^*-\lambda_MM^*-\lambda_HH^*-2bT(1-\frac{p-\mu_t}{2bT})
\\\nonumber&<&0; ~\mbox{hence diagonal dominance and conclusion of the proof.} \end{eqnarray}

From the proposition, the following conclusion is made:  \begin{lem}The endemic equilibrium is locally asymptotically stable if $R_i^2=\frac{r\delta_{_M}(1-\epsilon)+\gamma_{_I}}{\mu_e+\delta_{_M}+\lambda_IT^*}$, that is, if the basic infection-induced rupture rate per erythrocyte with immune activation is less than the total mortality rate of the erythrocytes.\end{lem}
\section{Estimating the severity of the co-infection } 
Let the basic acting factors of a co-infection of malaria and visceral helminthiasis be concentration of pathogenic multiplying antigens, $H(t)$, and $M(t)$, defined in the previous section, concentration of antibodies, $A(t)=A_{_M}+A_{_H}$, where $A_{_M}$ and $A_{_H}$ are the antibodies against the malaria parasites and helminths larvae respectively; concentration of plasma cells $C(t)$ and the relative characteristic of the affected organ, $m(t)$, in this case, red blood cell depletion. Consider the following model:-
\begin{eqnarray}\begin{array}{lll}
\frac{dM}{dt}&=&(\beta_{_M}-\gamma_{_M} A_M)M\\
\frac{dH}{dt}&=&(\beta_{_H}-\gamma_{_H} A_H)H\\
\frac{dC}{dt}&=&\alpha_{_M} A_MM+\alpha_{_H } A_HH-\mu_{_C}(C-C^o)\\
\frac{dA_M}{dt}&=&\rho_{_M }C-\mu_{_A} A_M -\eta_{_M}\gamma_{_M} MA_M\\
\frac{dA_H}{dt}&=&\rho_{_H } C-\mu_{_A} A_H -\eta_{_H}\gamma_{_H} HA_H\\
\frac{dm}{dt}&=&\sigma_{_M} M+\sigma_{_H} H -\mu_M m\end{array}
{\label{QN}}\end{eqnarray}
where $\beta_{_M}$ is the average rate of malaria parasite multiplication, and $\beta_{_H}$ is the average rate of helminths larvae invasion of red blood cells, $\gamma_{_H} A_H H$ and $\gamma_{_M} A_M M$ is the number of antigens neutralized by the respective antibodies during the time interval $\triangle t$. $\gamma_{_M}$ and $\gamma_{_H}$ are therefore parameters connected with the probability for an antigen to be neutralized after encountering the antibodies. The third equation describes the growth of plasma cells. This is the population of plasma cells, producers of antibodies; in this model, the antibodies are the substrates capable of binding the malaria parasites, and the helminths larvae, therefore, the number of lymphocytes stimulated in this way is proportional to $A_MM$ and $A_HH$. This is the relationship describing the increase of plasma cells over a normal level $C^o$, that is the constant level of plasma cells in a healthy human. $\alpha_{_M}$, $\alpha_{_H }$ are defined as the coefficients allowing for the probability of antigen-antibody collision respectively, the stimulation of cascade reaction, and the number of newly generated cells. We define $\frac{1}{\mu_{_C}}$ as the time it takes the plasma cells to age, implying that $\mu_{_C}(C-C^o)\triangle t$ is the decrease in the number of plasma cells due to aging \cite{March}. \\

Let $\rho_{_M }C$ and $\rho_{_H } C$ be the generation of antibodies fighting malaria parasites and helminths larvae by plasma cells, giving $\rho=\rho_{_M}+\rho_{_H}$ as the total rate of antibodies production by the plasma cells. The total number of eliminated antigens that were neutralized by antibodies is equal to $\gamma_{_M} A_MM +\gamma_{_H} A_HH$, so if neutralization of one antigen requires $\eta_{_M}+\eta_{_H}$ antibodies, then this implies that $\eta_{_M} \gamma_{_M} A_MM+\eta_{_H}\gamma_{_H} A_HH $ describes the total decrease in the number of antibodies due to binding with malaria parasites and helminths larvae. The total drop in antibodies population due to aging, where $\mu_{_A}$ is the per capita rate of decay of antibodies, is given by $\mu_{_A} A_M+\mu_{_A} A_H$. \\

Next we consider depletion of the red blood cells during the co-infection of malaria and helminths, due to the decrease in the activity of organs providing the delivery of immunologic materials, such as leukocytes, lymphocytes, and antibodies, that are necessary for the struggle with multiplying antigens. Assuming that the productivity of the red blood cells depend on the degree of invasion by malaria parasites and helminths larvae. Let $R$ be the relative characteristic of a mass of healthy red blood cells and $R^d$ the corresponding characteristic of the healthy red blood cells in an invaded red blood cell mass. Define $m$ as $1-\frac{R^d}{R}$, that is, $m$ is relative characteristic of damage to the red blood cell mass. For non invaded red blood cells, $m$ is equal to zero naturally, and for completely impregnated red blood cells, $m$ equals $1$. $\sigma_{_M} M+\sigma_{_H} H$ signifies the total degree of depletion of red blood cells. It is assumed that the increase in relative depletion of red blood cells is proportional to the number of antigens $M+H$, described by $\sigma_{_M} M+\sigma_{_H} H $, where $\sigma_{_M}, \sigma_{_H}$ are constants. A decrease in damage is caused by the recovery activity of an organism, that is, $\mu_M$ is the restoration of the red blood cells.  
\subsection{Steady states during the severity of the co-infection}
For stationary solutions, we equate the derivatives to zero and solve with initial conditions. Note that at $t=0$, $C=C^o$,~ $A^*_M=A_M$,~ $A^*_H=A_H$ are the values for a non invaded red blood cell mass for $M = 0 = H$. This implies that the trivial solution, describing the healthy state of red blood cells is \begin{eqnarray}M=0,~H=0,~ C=C^o,~A^*_M=\frac{\rho_{_M }C^o}{\mu_{_A}},A_H^*=\frac{\rho_{_H } C^o}{\mu_{_A}}, ~m=0{\label{EQ}}\end{eqnarray} Thus we have the following conclusion: \begin{lem}The concentration of antigen population and damaged mass of red blood cells at disease-free state are equal to zero, and the quantities of plasma cells, $C$ and antibodies $A_{_M}$ and $A_{_H}$ correspond to the values of an immunological status of a healthy individual.\end{lem} Consider the system of points $(M,H,C,A_M,A_H,m)$ in the neighborhood of the equilibrium point \\$(M^*,H^*,C^*,A^*_M,A_H^*,m)$. Let $M(t)=u(t),~H(t)=v(t),~C(t)=C^*+w(t),A_M(t)=A^*_M+x(t),A_H(t)=A^*_H+y(t),~m(t)=z(t)$ where $u,~v,~w,~x~,y,~z$ are small deviations of functions from this equilibrium state (\ref{EQ}). Substituting these expressions into the system of equations (\ref{QN}) and assuming that the deviations are small, neglecting small variables of the second order of smallness we obtain:
\begin{eqnarray}\begin{array}{l}
\frac{du}{dt}=(\beta_{_M}-\gamma A^*_M)u\\
\frac{dv}{dt}=(\beta_{_H}-\gamma A^*_H)v\\
\frac{dw}{dt}=\alpha_{_M} A^*_Mu+\alpha_{_H } A^*_Mv-\mu_{_C} w\\
\frac{dx}{dt}=\rho_{_M }w-\eta_{_M}\gamma_{_M} A^*_Mu-\mu_{_A} x\\
\frac{dy}{dt}=\rho_{_H } w-\eta_{_H}\gamma_{_H} A^*_Hv-\mu_{_A} y\\
\frac{dz}{dt}=\sigma_{_M} u+\sigma_{_H} v-\mu_M z\end{array}
\end{eqnarray} The corresponding Jacobian \begin{eqnarray}\frac{d{J}}{dt}=\left(\begin{array}{llllll}\beta_{_M}-\gamma_{_M} A^*_M&0&0&0&0&0\\0&\beta_{_H}-\gamma_{_H}A^*_H&0&0&0&0\\\alpha_{_M}A_M&\alpha_{_H}A_H&-\mu_{_C}&0&0&0\\-\eta_{_M}\gamma_{_M} A^*_M &0&\rho_{_M }&-\mu_{_A} &0&0\\0 &-\eta_{_H}\gamma_{_H} A^*_H&\rho_{_H } &0&-\mu_{_A} &0\\ \sigma_{_M}&\sigma_{_H} &0&0&0&-\mu_M\end{array}\right){J}\end{eqnarray} where ${J}=(u,v,w,x,y,z)^T(t)$, with distinct eigenvalues $-\mu_M,-\mu_{_A},-\mu_{_A},-\mu_{_C}$,~ giving the following result:  \begin{lem}All small deviations from stationary solution (\ref{EQ}) for $\beta_{_M} < \gamma_{_M} A^*_M$ and $\beta_{_H} < \gamma_{_H} A^*_H$ tend to zero with time, that is, the total number of malaria parasites multiplication, plus the total number of helminths larvae that invade the red blood cells at a given time must be less than the total number of malaria parasites and helminths larvae neutralized by the antibodies during an episode when both malaria parasites and helminths larvae are present in the red blood cells. This implies asymptotic stability of the invasion-free state.\end{lem}This leads to the following conclusion:  \begin{lem}From (\ref{QN}), at steady state, the small infection dose that does not lead to the loss of stability is estimated by \begin{eqnarray}\begin{array}{ll}0<M_o<\frac{\mu_{_A}(\gamma_{_M}A^*_{_M}-\beta_{_M})}{\beta_{_M}\gamma_{_M}\eta_{_M}}=M^*,&0<H_o<\frac{\mu_{_A}(\gamma_{_H}A^*_{_H}-\beta_{_H})}{\beta_{_H}\gamma_{_H}\eta_{_H}}\end{array}=H^*\end{eqnarray} where $M^*$ and $H^*$ are the immunological barrier values for malaria and helminthiasis respectively.\end{lem}
The immune barrier is exceeded when an infection dose $M_o$ and $H_o$ satisfies conditions $M_o > M^*$ and $H_o >H^*$, and is not exceeded otherwise. Biologically, it implies that, if in the case of red blood cell invasion by a small dose of malaria parasites and helminths larvae, the immunological barrier cannot be exceeded, then regardless of the infectious dose, the invasion does not develop, that is, the number of malaria parasites and helminths larvae in the organism decreases with time tending to zero, and the red blood cells are restored. In addition, the elevation of $C^o$, that is of the level of immunocompetent cells in a healthy red blood cells mass increases the immunological barrier, since $A^*_M=\frac{\rho_{_M }C^o_{_M}}{\mu_{_A}}$ and $A^*_H=\frac{\rho_{_H } C^o_{_H}}{\mu_{_A}}$ and therefore, it is an effective method of prophylaxis, and possibly, of disease treatment. This leads to the following lemma.  \begin{lem} The immunocompetent cells threshold values against malaria and helminths are given by \begin{eqnarray}\begin{array}{ll}0<M_o<\frac{\gamma_{_M}\rho_{_M}C^o_{_M}-1}{\beta_{_M}^2\gamma_{_M}\eta_{_M}}=M^*,&0<H_o<\frac{\gamma_{_H}\rho_{_H}C^o_{_H}-1}{\beta_{_H}^2\gamma_{_H}\eta_{_H}}\end{array}=H^*\end{eqnarray} with $C^o_{_M}=\frac{1}{\gamma_{_M}\rho_{_M}}$ and $C^o_{_H}=\frac{1}{\gamma_{_H}\rho_{_H}}$ as the threshold values of the immunocompetent cells respectively, below which, no immune response is expected. \end{lem}

\subsection{Results of the severity of the co-infection}
Chronic form of either disease is caused by an insufficiently effective reaction of the immune system that happens when $\alpha_{_M}$ or $\alpha_{_H}$ is small, that is if the immune system's reaction is weak. In this case, $M\rightarrow M_{max}$ and $H\rightarrow H_{max}$; and on the other hand, $A^*_{M}$ and $A^*_{_H}$ drop respectively. The outcome of the co-infection depends on the derivatives $\frac{dM}{dt}$ and $\frac{dH}{dt}$, if they can become negative and how long they do not change signs. It is noted that $\frac{dM}{dt} <0$ if $M(t)>0$ and $A_{_M}(t)>\frac{\beta_{_M}}{\gamma_{_M}}$ and $\frac{dH}{dt} <0$ if $H(t)>0$ and $A_{_H}(t)>\frac{\beta_{_H}}{\gamma_{_H}}$. Thus we have the following conclusion:  \begin{lem} If we assume that infection of either or both malaria and visceral helminthiasis has happened, then the $A_{_M}(t)>\frac{\beta_{_M}}{\gamma_{_M}}$ and/or $A_{_H}(t)>\frac{\beta_{_H}}{\gamma_{_H}}$ is the necessary and sufficient condition for $\frac{dM}{dt}$ and $\frac{dH}{dt}$ to be negative. \end{lem}
If the immunological barrier is not to be exceeded, $(M_o < M)$ and $(H_o < H)$, then only {\bf{CASE I }}of the following four cases is possible:-
\begin{description}
\item[CASE I]$\frac{dM}{dt} <0, \frac{dH}{dt} <0$ in a large interval of time. In this case, there will be a {\it{subclinical form of the co-infection}}. This happens since $\beta_{_M} < \gamma_{_M}A^*_{M}$  and $\beta_{_H} < \gamma_{_H}A^*_{H}$ in which case either there is an effective (normal) immune response, when $\alpha_{_M}(\rho_{_M}+\rho_{_H}) >\mu_{_C}\eta_{_M}\gamma_{_M}$ and $\alpha_{_H}(\rho_{_H}+\rho_{_H}) >\mu_{_C}\eta_{_H}\gamma_{_H}$,  and weak response, (immunodeficiency) when $\alpha_{_M}(\rho_{_M}+\rho_{_H}) <\mu_{_C}\eta_{_M}\gamma_{_M}$ and $\alpha_{_H}(\rho_{_H}+\rho_{_H}) < \mu_{_C}\eta_{_H}\gamma_{_H}$. That is, when the infection dose is small, lower than the immunological barrier values, the removal of the malaria parasite or helminths larvae from the red blood circulation depends neither on the doses of infections nor on the strength of immune response. The removal of either antigen is provided by the antibody levels $A^*_{_M}$ or $A^*_{_H}$ against each disease that are present in the body at the time of infection. When the doses of malaria parasites and/or visceral helminthiasis increase to considerable extents as compared to the immune barrier, the strength of the immune response begins to play an important role.\\

In case of $\beta_{_M} > \gamma_{_M}A^*_{_M},~ \beta_{_H} > \gamma_{_H}A^*_{_H},~ \frac{dM}{dt} >0, ~\frac{dH}{dt} > 0$ for an interval $(t_1,t_2)$, and assuming that $M\rightarrow M_{max}$ and $H\rightarrow H_{max}$ at a point $t_1$ in the interval, and then decrease to $M\rightarrow M_{min}$ and $H\rightarrow H_{min}$, we have two cases: \\EITHER 
\item[CASE II] $\frac{dM}{dt} <0, \frac{dH}{dt} <0$ within a sufficiently large time interval $(t_1,t_2)$ where $t_2$ is the time when $M\rightarrow M_{min}$ and $H\rightarrow H_{min}$. Hence, {\it{acute form of the co-infection}} is obtained in which case, $\beta_{_M}>\gamma_{_M}\eta_{_M},~\beta_{_H}>\gamma_{_H}\eta_{_H}$ and therefore the immunological barrier against malaria parasites or helminths larvae does not exist. Therefore, there would be a rapid increase of malaria parasites and helminths larvae in the red blood cells up to the values exceeding the infection doses, and then rapidly get eliminated due to a strong and effective immune response that leads to the production of antibodies against the malaria parasites and helminths larvae in quantities sufficient for the elimination of both infections. Note that the higher $\beta_{_M}$ and $\beta_{_H}$ are, the higher $A^*_M$ and $A^*_H$ respectively; hence the faster the maximum value of malaria and helminths fighting antigens quantities are attained and the faster the process will stop. This gives the following result:  \begin{lem}With a high rate of malaria parasite multiplication, or a high rate of red blood invasion by the helminths larvae, or more generally, a high dose of infection by either disease, with other parameters constant, the amount of malaria parasites and helminths larvae that stimulate the immune system effectively is attained more quickly.\end{lem} This implies that $M_{max}$ or $H_{max}$ does not depend on the dose of infection but is determined by the immune status in relation to the malaria parasites or helminths larvae; that is, by the set of model parameters. The dose of the infection affects the time when $M_{max}$ or $H_{max}$ is reached. In this form of infection, there is red blood depletion due to malaria, $\sigma_{_M}$ and due to helminths, $\sigma_{_H}$.    \\OR
\item[CASE III] $\frac{dM}{dt} <0, \frac{dH}{dt} <0$ within a sufficiently small time interval $(t_1,t_2)$, then the {\it{chronic form of the co-infection}} is expected. \\If however a point $t_1$ does not exist, that is if, 
\item[CASE IV]$\frac{dM}{dt} > 0, \frac{dH}{dt} >0$ within an infinitely large interval of time, then {\it{a lethal outcome of the co-infection}} is obtained. This happens when the formation of plasma cells is delayed and in turn a delay in the production of antibodies specific to malaria parasites or helminths larvae. This leads to the following result:  \begin{lem}In order to prevent a hyper-toxic form of the co-infection of malaria and helminthiasis, it is necessary to delay the replication rate of malaria parasites $\beta_{_M}$ and the invasion rate of helminths larvae, $\beta_{_H}$. \end{lem}
\end{description}

\subsection{Threshold values for the severity of the co-infection}
If both or one of the antigens is eliminated, then from (\ref{QN}) at steady state, the following equilibria exist: 
\begin{eqnarray}\begin{array}{ll}E_o&=(0,0,C,\frac{\rho_{_M }C}{\mu_{_A}},\frac{\rho_{_H } C}{\mu_{_A}},0)~~~\mbox{the disease-free,}\\E_H&=\left[0,\frac{\mu_{_A}\mu_{_C}}{(\alpha_{_H }\rho_{_H }-\mu_{_C}\gamma_{_H}\eta_{_H})},\frac{\alpha_{_H }\beta_{_H}\mu_{_A}}{\gamma_{_H}(\alpha_{_H }\rho_{_H }-\mu_{_C}\gamma_{_H}\eta_{_H})},\frac{\rho_{_M}\alpha_{_H }\beta_{_H}\mu_{_A}}{\mu_{_A}\gamma_{_H}(\alpha_{_H }\rho_{_H }-\mu_{_C}\gamma_{_H}\eta_{_H})},\frac{\beta_{_H}}{\gamma_{_H}}\right],\\&\left[\frac{\sigma_{_H}\mu_{_C}\mu_{_A}}{\mu_M(\alpha_{_H }\rho_{_H }-\mu_{_C}\gamma_{_H}\eta_{_H})}\right]\\E_M&=\left[\frac{\mu_{_A}\mu_{_C}}{(\alpha_{_M}\rho_{_M}-\mu_{_C}\gamma_{_M}\eta_{_M})},0,\frac{\alpha_{_M}\beta_{_M}\mu_{_A}}{\gamma_{_M}(\alpha_{_M}\rho_{_M}-\mu_{_C}\gamma_{_M}\eta_{_M})},\frac{\beta_{_M}}{\gamma_{_M}},\frac{\rho_{_H }\alpha_{_M}\beta_{_M}\mu_{_A}}{\mu_{_A}\gamma_{_M}(\alpha_{_M}\rho_{_M}-\mu_{_C}\gamma_{_M}\eta_{_M})}\right],\\&\left[\frac{\sigma_{_M}\mu_{_A}\mu_{_C}}{\mu_M(\alpha_{_M}\rho_{_M}-\mu_{_C}\gamma_{_M}\eta_{_M})}\right]
\end{array}\end{eqnarray} where endemicity of either equilibrium depends on the respective threshold values.\\
For competitive dominance of helminths larvae by the malaria parasites, the following equilibrium point will be attained : \begin{eqnarray}\begin{array}{llllll}E_{M_H}=\frac{\mu_{_A}}{\eta_{_M}\gamma_{_M}(R^2_{_{eM}}-1)},\frac{\mu_{_A}\mu_{_C}}{\alpha_{_H }(\rho_{_M}+\rho_{_H })(1-R^2_{_{oH}})},\frac{\mu_{_A}[\alpha_{_M}(\rho_{_M}+\rho_{_H })+\mu_{_C}\eta_{_M}(\beta_{_M}-\gamma_{_M})]}{\mu_{_C}\eta_{_M}\gamma_{_M}\rho_{_M}(R^2_{_{eM}}-1)},\\\frac{\beta_{_M}}{\gamma_{_M}},\frac{\beta_{_H}}{\gamma_{_H}},\frac{\sigma_{_M}\mu_{_A}}{\mu_M\eta_{_M}\gamma_{_M}(R^2_{_{eM}}-1)}\end{array}\end{eqnarray} with $R^2_{_{eM}}>1 > R^2_{_{Ho}}$ where $R_{_{eM}}$ is the basic reproductive potential for a successful malaria dominance in presence of helminths and \begin{eqnarray}\begin{array}{ll}R^2_{_{eM}}=\frac{\alpha_{_M}(\rho_{_M}+\rho_{_H })}{\mu_{_C}\eta_{_M}\gamma_{_M}},&R^2_{_{oH}}=\frac{\mu_{_C}\eta_{_H}\gamma_{_H}}{\alpha_{_H }(\rho_{_M}+\rho_{_H })}\end{array}\end{eqnarray}
Thus, we obtain the following result: \begin{lem}For competitive exclusion of visceral helminthiasis by malaria, $R^2_{_{eM}}>1 > R^2_{_{Ho}}$. That is, the total rate if malaria parasites-antibody collision must be greater than the loss due to neutralizing by the antibodies; in this case, the malaria parasites have become resilient in their quest for survival.\end{lem}
If helminths dominates malaria, then the endemic equilibrium will be given by \begin{eqnarray}\begin{array}{llllll}E_{H_M}=\frac{\mu_{_A}\mu_{_C}}{\alpha_{_M}(\rho_{_M}+\rho_{_H })(1-R^2_{_{oM}})},\frac{\mu_{_A}}{\eta_{_H}\gamma_{_H}(R^2_{_{eH}}-1)},\frac{\mu_{_A}[\alpha_{_H }(\rho_{_M}+\rho_{_H })+\mu_{_C}\eta_{_H}(\beta_{_H}-\gamma_{_H})]}{\rho_{_H }\mu_{_C}\eta_{_H}\gamma_{_H}(R^2_{_{eH}}-1)},\frac{\beta_{_M}}{\gamma_{_M}},\\
\frac{\beta_{_H}}{\gamma_{_H}},\frac{\sigma_{_H}\mu_{_A}}{\mu_M\eta_{_H}\gamma_{_H}(R^2_{_{eH}}-1)}\end{array}\end{eqnarray} which exists whenever $R^2_{_{eH}} > 1> R_{_{oM}}^2$ where $R_{_{eH}}$ is the basic within-host invasion potential of helminths larvae in presence of malaria parasites and 
\begin{eqnarray}\begin{array}{ll}R^2_{_{eH}}=\frac{\alpha_{_H }(\rho_{_M}+\rho_{_H })}{\mu_{_C}\eta_{_H}\gamma_{_H}}&R^2_{_{oM}}=\frac{\mu_{_C}\eta_{_M}\gamma_{_M}}{\alpha_{_M}(\rho_{_M}+\rho_{_H })}\end{array}\end{eqnarray}
 This gives the following similar result:
 \begin{lem}If visceral helminthiasis dominates malaria, $R^2_{_{eH}}>1 > R^2_{_{oM}}$. Similarly, the helminths larvae are resilient but the malaria parasites are being neutralized by the antibodies, and so $ R^2_{_{oM}} < 1$.   \end{lem}
However, if both antigens compete for coexistence, then one of the following endemic equilibria is expected:
\begin{description}
\item[A.] {\bf{Case of weak sub thresholds}} \begin{enumerate}\item[CASE I] ~~~$R_{oM}<1,~R_{eH}>1,~R_{eMH}>1$\begin{eqnarray}\nonumber\begin{array}{llllll}E^c_{H_M}=&&\frac{\mu_{_A}\mu_{_C}}{\alpha_{_M}(\rho_{_M}+\rho_{_H })(1-R_{_{oM}}^2)},\frac{\mu_{_A}}{\eta_{_H}\gamma_{_H}(R^2_{_{eH}}-1)},\\&&\left(\frac{\mu_{_A}[\alpha_{_M}(\rho_{_M}+\rho_{_H })+\mu_{_C}\eta_{_M}(\beta_{_M}-\gamma_{_M})]}{\rho_{_M}\alpha_{_M}(\rho_{_M}+\rho_{_H })(1-R_{_{oM}}^2)}+\frac{\mu_{_A}[\alpha_{_H }(\rho_{_M}+\rho_{_H })+\mu_{_C}\eta_{_H}(\beta_{_H}-\gamma_{_H})]}{\rho_{_H }\mu_{_C}\eta_{_H}\gamma_{_H}(R^2_{_{eH}})-1}\right),\\&&\frac{\beta_{_M}}{\gamma_{_M}},\frac{\beta_{_H}}{\gamma_{_H}},\frac{\left[\sigma_{_M}\mu_{_A}\mu_{_C}^2\eta_{_M}\gamma_{_M}\alpha_{_M}(\rho_{_M}+\rho_{_H })(1-R_{_{oM}}^2)+\sigma_{_H}\mu_{_A}\mu_{_C}^2\eta_{_H}\gamma_{_H}(R^2_{_{eH}}-1)\right]}{\mu_M(\alpha_{_M}\alpha_{_H }(\rho_{_M}+\rho_{_H })^2+\mu_{_C}^2\eta_{_M}\eta_{_H}\gamma_{_M}\gamma_{_H})(R^2_{_{eMH}}-1)}\end{array}\\\end{eqnarray}
\item[CASE II] ~~~$R_{eM}>1,~R_{oH}<1,~R_{eMH}>1$
\begin{eqnarray}\nonumber\begin{array}{llllll}E^c_{M_H}=&&\frac{\mu_{_A}}{\eta_{_M}\gamma_{_M}(R_{_{eM}}^2-1)},\frac{\mu_{_A}\mu_{_C}}{\alpha_{_H }(\rho_{_M}+\rho_{_H })(1-R^2_{_{oH}})},\\&&\left(\frac{\mu_{_A}[\alpha_{_M}(\rho_{_M}+\rho_{_H })+\mu_{_C}\eta_{_M}(\beta_{_M}-\gamma_{_M})]}{\rho_{_M}\mu_{_C}\eta_{_M}\gamma_{_M}(R_{_{eM}}^2-1)}+\frac{\mu_{_A}[\alpha_{_H }(\rho_{_M}+\rho_{_H })+\mu_{_C}\eta_{_H}(\beta_{_H}-\gamma_{_H})]}{\rho_{_H }\alpha_{_H }(\rho_{_M}+\rho_{_H })(1-R^2_{_{oH}})}\right),\\&&\frac{\beta_{_M}}{\gamma_{_M}},\frac{\beta_{_H}}{\gamma_{_H}},\frac{\left[\sigma_{_M}\mu_{_A}\mu_{_C}^2\eta_{_M}\gamma_{_M}(R_{_{eM}}^2-1)+\sigma_{_H}\mu_{_A}\mu_{_C}^2\eta_{_H}\gamma_{_H}(1-R^2_{_{oH}})\right]}{\mu_M(\alpha_{_M}\alpha_{_H }(\rho_{_M}+\rho_{_H })^2+\mu_{_C}^2\eta_{_M}\eta_{_H}\gamma_{_M}\gamma_{_H})(R^2_{_{eMH}}-1)}
\end{array}\\\end{eqnarray}
\end{enumerate}
\item[B.] {\bf{Case of strong sub thresholds}}~~~$R_{oM}<1,~R_{oH}<1,~R_{oMH}<1$
\begin{eqnarray}\nonumber\begin{array}{llllll}E^c_{HM}=&&\frac{\mu_{_A}\mu_{_C}}{\alpha_{_M }(\rho_{_M}+\rho_{_H })(1-R_{_{oM}}^2)},\frac{\mu_{_A}\mu_{_C}}{\alpha_{_H }(\rho_{_M}+\rho_{_H })(1-R^2_{_{oH}})},\\&&\left(\frac{\mu_{_A}[\alpha_{_M}(\rho_{_M}+\rho_{_H })+\mu_{_C}\eta_{_M}(\beta_{_M}-\gamma_{_M})]}{\rho_{_M}\alpha_{_M }(\rho_{_M}+\rho_{_H })(1-R_{_{oM}}^2)}+\frac{\mu_{_A}[\alpha_{_H }(\rho_{_M}+\rho_{_H })+\mu_{_C}\eta_{_H}(\beta_{_H}-\gamma_{_H})]}{\rho_{_H }\alpha_{_H }(\rho_{_M}+\rho_{_H })(1-R^2_{_{oH}})}\right),\\&&\frac{\beta_{_M}}{\gamma_{_M}},\frac{\beta_{_H}}{\gamma_{_H}},\frac{\left[\sigma_{_M}\mu_{_A}\mu_{_C}^2\eta_{_M}\gamma_{_M}\alpha_{_M }(\rho_{_M}+\rho_{_H })(1-R_{_{oM}}^2)+\sigma_{_H}\mu_{_A}\mu_{_C}^2\eta_{_H}\gamma_{_H}\alpha_{_H }(\rho_{_M}+\rho_{_H })(1-R^2_{_{oH}})\right]}{\mu_m\mu_{_C}(\rho_{_M}+\rho_{_H })(\alpha_{_M}\eta_{_H}\gamma_{_H}+\alpha_{_H }\eta_{_M}\gamma_{_M})(1-R^2_{_{oMH}})}\end{array}\\\end{eqnarray} \end{description}where 
$R^2_{_{eMH}}=\frac{\mu_{_C}(\rho_{_M}+\rho_{_H })[\alpha_{_M}\eta_{_H}\gamma_{_H}+\alpha_{_H }\gamma_{_M}\eta_{_M}]}{[\alpha_{_M}\alpha_{_H }(\rho_{_M}+\rho_{_H })^2+\mu_{_C}^2\eta_{_M}\eta_{_H}\gamma_{_M}\gamma_{_H}]} >1$ and \\$R^2_{_{oMH}}=\frac{[\alpha_{_M}\alpha_{_H }(\rho_{_M}+\rho_{_H })^2+\mu_{_C}^2\eta_{_M}\eta_{_H}\gamma_{_M}\gamma_{_H}]}
{\mu_{_C}(\rho_{_M}+\rho_{_H })[\alpha_{_M}\eta_{_H}\gamma_{_H}+\alpha_{_H }\gamma_{_M}\eta_{_M}]}<1$
 is the within-host basic reproductive potential for a successful co-infection of malaria and helminths. This gives the following result:
 \begin{lem}The malaria parasites and helminths larva will co-exist, for strong sub thresholds and one of them will dominate the other for weak sub thresholds. This implies that there may be competitor-oscillatory coexistence until the system bi-stabilizes to endemic equilibria, depending on the immune action against either or both antigens.\end{lem} Further from the above lemma, the following conclusion is made:
 \begin{lem}For endemicity of the within-host co-infection of malaria and visceral helminthiasis, one of the reproduction numbers must be greater than 1 and coexistence equilibria exists; no coexistence exists when both reproductive potentials are greater than 1.\end{lem}
\section{Discussion}
In the model for the co-infection of malaria and visceral helminthiasis, we have assumed that an individual gets malaria infection during invasion by helminth larvae. In absence of immune response, our results show that there is no disease-free state, but an endemic state is attained via damped oscillations towards a fixed point (see Figure \ref{1} A,C and B). Our results showed that antigens invade the blood system if the rate of red cell rupture per invading merozoite is greater than one as seen in Lemma \ref{lem1}. If more merozoite are released, there exists malaria-only endemic point. However, both antigens co-exist if the mean infection burden is greater than one. In this case, there is a threshold value for drug action below which no recovery of host is expected, Lemma \ref{lem4}. In presence of immune response, three equilibrium states exist. The initial invasion state, and secondly, the unstable state when the immune population has been triggered and the antigens have been eliminated. The third state represents the endemic state which is stable if the infection-induced rupture rate per erythrocyte with immune activation is less than the total mortality of the erythrocyte (see Lemma \ref{lem6} and Figure \ref{helfig2}). A model for severity of the co-infection shows in Figure \ref{helfig2}B that an immune response will be delayed until immunological barrier values for malaria and helminth are exceeded. Our results show further that once the immune response is triggered, the system in inherently unstable and will not return to zero, Lemma \ref{lem7}.\\

In this paper, a co-infection of within-host malaria and visceral helmnithiasis is studied. The interest was to study immune response and impairment when bot antigens invade the red blood cells. The process of immune response is more complicated than modelled here, but the model has important consequences for the success of disease control programs. This model is a conceptually simple and a logical extension of the classic individual species models that have formed the foundation of epidemiological theory over the last 30 years. As such, it should provide a valuable framework for understanding the dynamics of parasite co-infection with immune response and evaluating the wide-ranging implications of disease control strategies in the future. Although with constraints due to limited data, the study shows that the epidemiology and population dynamics of within-host-microparasite-macroparasite communities are highly complex, resulting in different predictions from those made by the corresponding single parasite species models. If better estimates of parameters become available, more detailed studies can be done by taking into account some of the aspects that have been overlooked.

\end{document}